\documentclass{aa}  

\usepackage{graphicx}
\usepackage{txfonts}
\usepackage[colorlinks=true, linkcolor=blue, citecolor=blue, urlcolor=blue]{hyperref}
\usepackage{graphics}
\usepackage{amsmath}
\usepackage{amssymb}
\usepackage{url}
\usepackage{xcolor}
\usepackage{physics}
\usepackage{booktabs}
\usepackage{orcidlink}
\usepackage{siunitx}
\usepackage{xspace}
\usepackage{subcaption}
\usepackage{threeparttable}
\usepackage[capitalise]{cleveref}

\makeatletter
\renewcommand*\aa@pageof{, page \thepage{} of \pageref*{LastPage}}
\makeatother

\newcommand{\Msun}{$\mathrm{M}_\odot$\xspace}
\newcommand{\msol}{\mathrm{M}_\odot\xspace}
\newcommand{\mstar}{\mathrm{M}_\star}
\newcommand{\logM}{$\log{(\mstar/\msol)}$\xspace}
\newcommand{\Msunpyr}{\,M$_{\odot}$yr$^{-1}$\xspace}

\newcommand{\lpok}{$\log{(P/k)}$\xspace}

\newcommand{\logOH}{12+log(O/H)\xspace}

\newcommand{\jwst}{\textit{JWST}\xspace}
\newcommand{\hst}{\textit{HST}\xspace}

\newcommand{\ha}{\ensuremath{\ion{H}{$\alpha$}}\xspace}
\newcommand{\hb}{\ensuremath{\ion{H}{$\beta$}}\xspace}

\newcommand{\hii}{\ion{H}{ii}\xspace}

\newcommand{\oiii}{\ensuremath{\left[\ion{O}{iii}\right]}\xspace}
\newcommand{\oiiilam}{\ensuremath{\left[\ion{O}{iii}\right]\,{\lambda 5007}}\xspace}
\newcommand{\oiiiablam}{\ensuremath{\left[\ion{O}{iii}\right]\,{\lambda 4959,5007}}\xspace}
\newcommand{\oii}{\ensuremath{\left[\ion{O}{ii}\right]}\xspace}
\newcommand{\oiilam}{\ensuremath{\left[\ion{O}{ii}\right]\,{\lambda 3727,3729}}\xspace}
\newcommand{\nii}{\ensuremath{\left[\ion{N}{ii}\right]}\xspace}
\newcommand{\sii}{\ensuremath{\left[\ion{S}{ii}\right]}\xspace}
\newcommand{\siilam}{\ensuremath{\left[\ion{S}{ii}\right]\,{\lambda 6717,6731}}\xspace}
\newcommand{\niilam}{\ensuremath{\left[\ion{N}{ii}\right]\,{\lambda 6548,6584}}\xspace}

\newcommand{\halpha}{H$\alpha$\xspace}
\newcommand{\hbeta}{H$\beta$\xspace}
\newcommand{\neiii}{\ensuremath{\left[\ion{Ne}{iii}\right]}\xspace}
\newcommand{\neiiilam}{\ensuremath{\left[\ion{Ne}{iii}\right]\,{\lambda 3867}}\xspace}
\newcommand{\heilam}{\ensuremath{\ion{He}{i}\,{\lambda 3889}}\xspace}
\newcommand{\ohno}{\textit{OHNO}\xspace}

\usepackage[normalem]{ulem}

\newcommand{\aref}[1]{\hyperref[#1]{Appendix~\ref{#1}}}

\defcitealias{Rigby:2011aa}{R11}
\defcitealias{Rigby:2018aa}{R18a}
\defcitealias{Rigby:2018ab}{R18b}
\defcitealias{Kewley:2002fk}{KD02}
\defcitealias{Whitaker:2014aa}{W14}
\defcitealias{Blanc:2015aa}{B15}
\defcitealias{Pettini:2004qe}{PP04}
\defcitealias{Shi:2007aa}{S07}
\defcitealias{Dopita:2016aa}{D16}
\defcitealias{Kobulnicky:2004aa}{KK04}
\defcitealias{Izotov:2006ab}{I06}
\defcitealias{Garnett:1995aa}{G95a}
\defcitealias{Garnett:1995ab}{G95b}
\defcitealias{Levesque:2014aa}{LR14}
\defcitealias{Kewley:2008aa}{KE08}
\defcitealias{Bian:2018aa}{BKD18}
\defcitealias{Sanders:2016aa}{S16}
\defcitealias{Strom:2018aa}{S18}
\defcitealias{Kewley:2019aa}{K19a}
\defcitealias{Kewley:2019ab}{K19b}
\defcitealias{Goldbaum:2016aa}{G16}
\defcitealias{Carton:2017aa}{C17}
\defcitealias{Acharyya:2019aa}{A19}
\defcitealias{Acharyya:2020aa}{A20}
\defcitealias{Acharyya:2020ab}{A20b}
\defcitealias{Sanchez:2018ab}{S18}
\defcitealias{Wylezalek:2018aa}{W18}
\defcitealias{Wang:2022aa}{W22}
\defcitealias{Feuillet:2024aa}{F24}
\defcitealias{Backhaus:2022aa}{B22}
\defcitealias{Curti:2020ab}{C20}
\defcitealias{Cataldi:2025aa}{C25}

\newcommand{\logU}{$\log{(U)}$\xspace}
\newcommand{\curti}{\citetalias{Curti:2020ab}\xspace}
\newcommand{\cataldi}{\citetalias{Cataldi:2025aa}\xspace}
\newcommand{\grizli}{\ensuremath{\textsc{grizli}}\xspace}
\newcommand{\edit}[1]{#1}
\newcommand{\newedit}[1]{#1}

\crefname{section}{Sect.}{Sects.}
\crefname{subsection}{Sect.}{Sects.}
\crefname{subsubsection}{Sect.}{Sects.}

\begin{document} 

\title{Spatially resolved gas-phase metallicity at $z \sim 2-3$ with \jwst/NIRISS}

\author{
Ayan Acharyya
\inst{1}\orcidlink{0000-0003-4804-7142}
\and
Peter J. Watson
\inst{1}\orcidlink{0000-0003-3108-0624}
\and
Benedetta Vulcani
\inst{1}\orcidlink{0000-0003-0980-1499}
\and
Tommaso Treu
\inst{2}\orcidlink{0000-0002-8460-0390}
\and
Kalina V. Nedkova
\inst{3,4}\orcidlink{0000-0001-5294-8002}
\and
Andrew J.\ Bunker
\inst{5}\orcidlink{0000-0002-8651-9879}
\and
Vihang Mehta
\inst{6}\orcidlink{0000-0001-7166-6035}
\and
Hakim Atek
\inst{7}\orcidlink{0000-0002-7570-0824}
\and
Andrew J. Battisti
\inst{8,9,10}\orcidlink{0000-0003-4569-2285}
\and
Farhanul Hasan
\inst{4}\orcidlink{0000-0002-0072-0281}
\and
Matthew J. Hayes
\inst{11}\orcidlink{0000-0001-8587-218X}
\and
Mason Huberty
\inst{12}\orcidlink{0009-0002-9932-4461}
\and
Tucker Jones
\inst{13}\orcidlink{0000-0001-5860-3419}
\and
Nicha Leethochawalit
\inst{14}\orcidlink{0000-0003-4570-3159}
\and
Yu-Heng Lin
\inst{6}\orcidlink{0000-0001-8792-3091}
\and
Matthew A. Malkan
\inst{2}\orcidlink{0000-0001-6919-1237}
\and
Benjamin Metha
\inst{15}
\and
Themiya Nanayakkara
\inst{16}
\and
Marc Rafelski
\inst{3,4}\orcidlink{0000-0002-9946-4731}
\and
Zahra Sattari
\inst{6}\orcidlink{0000-0002-0364-1159}
\and
Claudia Scarlata
\inst{12}\orcidlink{0000-0002-9136-8876}
\and
Xin Wang
\inst{17, 18, 19}\orcidlink{0000-0002-9373-3865}
\and
Caitlin M. Casey
\inst{20, 21}\orcidlink{0000-0002-0930-6466}
\and
Andrea Grazian
\inst{1}
\and
Anton M. Koekemoer
\inst{4}\orcidlink{0000-0002-6610-2048}
\and
Mario Radovich
\inst{1}
\and
Giulia Rodighiero
\inst{1, 22}
}

\institute{
INAF -- Osservatorio Astronomico di Padova, Vicolo Osservatorio 5, 35122 Padova, Italy
\email{ayan.acharyya@inaf.it}
\and
University of California, Los Angeles, Department of Physics and Astronomy, 430 Portola Plaza, Los Angeles, CA 90095, USA
\and
The Johns Hopkins University, Department of Physics and Astronomy, 3400 N. Charles Street, Baltimore, MD 21218, USA
\and
Space Telescope Science Institute, 3700 San Martin Drive, Baltimore, MD 21218, USA
\and
University of Oxford, Department of Physics, Keble Road, Oxford OX1 3RH, UK
\and
IPAC, California Institute of Technology, 1200 E. California Blvd, Pasadena, CA 91125, USA
\and  
Institut d’Astrophysique de Paris, CNRS, Sorbonne Universit\'e, 98bis Boulevard Arago, 75014, Paris, France
\and
International Centre for Radio Astronomy Research (ICRAR), University of Western Australia, M468, 35 Stirling Highway, Crawley, WA 6009, Australia
\and
Australian National University, Research School of Astronomy and Astrophysics, Canberra, ACT 2611, Australia
\and
ARC Centre of Excellence for All Sky Astrophysics in 3 Dimensions (ASTRO 3D), Australia
\and
Stockholm University, Department of Astronomy, AlbaNova University Center, SE-106 91 Stockholm, Sweden
\and
Minnesota Institute for Astrophysics, University of Minnesota, 116 Church Street SE, Minneapolis, MN 55455, USA
\and
Department of Physics and Astronomy, University of California, Davis, 1 Shields Ave, Davis, CA 95616, USA
\and
National Astronomical Research Institute of Thailand (NARIT), Mae Rim, Chiang Mai, 50180, Thailand
\and
School of Physics, The University of Melbourne, VIC 3010, Australia
\and
Centre for Astrophysics and Supercomputing, Swinburne University of Technology, PO Box 218, Hawthorn, VIC 3122, Australia
\and
School of Astronomy and Space Science, University of Chinese Academy of Sciences (UCAS), Beijing 100049, China
\and
National Astronomical Observatories, Chinese Academy of Sciences, Beijing 100101, China
\and
Institute for Frontiers in Astronomy and Astrophysics, Beijing Normal University, Beijing 102206, China
\and
Department of Physics, University of California, Santa Barbara, Santa Barbara, CA 93106, USA
\and
Cosmic Dawn Center (DAWN), Denmark
\and
Department of Physics and Astronomy, Università degli Studi di Padova, Vicolo dell’Osservatorio 3, I-35122 Padova, Italy
}

\authorrunning{A Acharyya et al.}
\titlerunning{PASSAGE: spatially resolved metallicities}

\date{Received ???? ??, ????; accepted ???? ??, ????}

\abstract
{
Spatially resolved gas-phase metallicity maps are a crucial element in understanding the chemical evolution of galaxies. We present spatially resolved metallicity maps obtained from NIRISS/WFSS observations. This is the first work presenting such analysis from slitless spectroscopy of multiple galaxies. We investigate the sources of ionization, metallicity and their relation to star-formation in a spatially-resolved sense for a sample of eight galaxies---four from JWST-PASSAGE and four from GLASS-JWST ERS. All but one galaxy are in the redshift range $1.9 \leq z \leq 2$, the outlier being at $z = 3.1$. Our sample covers a range of $8.0 <$ \logM $< 9.5$ in stellar mass, $0.2 <$ $\log{\rm{(SFR}}$/\Msunpyr) $< 1.1$ in star-formation rate (SFR) and $7.8 <$ \logOH $< 9.0$ in global gas metallicity. To resolve the question of SF-AGN separation in the absence of resolved \halpha+[NII] lines, we present a new SF-demarcation line in the \ohno parameter space, based on MAPPINGS v5.1 publicly available \hii region and AGN model grids, \newedit{with the caveat that these grids partially overlap the \ohno parameter space}. We present the mass-metallicity gradient relation for our sample, which show no clear trend with stellar mass. This might be because the high-$z$ galaxies have not yet started their accretion-dominated evolutionary phase. By interpreting the correlation between spatially resolved metallicity and SFR maps as a proxy for effective timescales of metal-transport in galaxies, we find a possible trend for this timescale to increase with stellar mass. This is consistent with the picture of more effective feedback in lower mass galaxies, and that massive galaxies generally have longer characteristic timescales.}

\maketitle

\section{Introduction}
\label{sec:intro}

The spatial distribution of heavy elements in the interstellar gas, typically measured as the abundance of Oxygen relative to Hydrogen, is a crucial diagnostic for the evolution of galaxies. While global measurements of galaxy metallicity provide indispensable clues about how galaxies evolve as a population \citep[e.g.,][and references therein]{Henry:1999aa, Kunth:2000aa, Tremonti:2004aa, Zahid:2011aa, Maiolino:2019aa, Thomas:2019aa, Kewley:2019ab}, spatially resolved metallicity studies offer insight into the local processes within each galaxy that drive said evolution. Spatially resolved metallicity studies are potential diagnostics of baryon flows within and around galaxies \citep[e.g.,][]{Rosales-Ortega:2012aa, Jones:2015aa, Wang:2017aa, Belfiore:2017aa, Thorp:2018aa, Belfiore:2019aa, Maiolino:2019aa, Venturi:2024aa, Lam:2026aa}.

Many recent studies have measured spatially resolved chemical abundances at low redshift ($z \lesssim 0.5$), thanks to multiple ground-based integral field spectroscopy (IFS) surveys \citep[e.g.,][]{Croom:2012aa, Bundy:2015aa, Sanchez:2012aa, Wisnioski:2015aa, Stott:2016aa, Poetrodjojo:2019aa}. Similar efforts were made in the intermediate-$z$ ($0.5 \lesssim z \lesssim 2.5$) Universe with the Hubble Space Telescope (\hst) \citep{Jones:2015aa, Wang:2017aa, Simons:2021aa} and with the James Webb Space Telescope (\jwst) \citep[e.g.,][]{, Venturi:2024aa, Barisic:2025aa, Ju:2025aa, Estrada-Carpenter:2025aa, Benotto:2026aa}. 

 \jwst has been transforming the field with its unprecedented sensitivity and angular resolution, allowing for large spatially resolved surveys of galaxies at high-$z$. In particular, the Wide Field Slitless Spectroscopy (WFSS) mode  provides spatially resolved images and spectra of several hundred galaxies in a single pointing. This capability is available with the Near Infrared Camera (NIRCam) and Near Infrared Slitless Spectrograph (NIRISS). Surveys such as FRESCO \citep{Oesch:2023aa}, JADES \citep{Rieke:2023aa, Sun:2024aa}, NGDEEP \citep{Pirzkal:2024aa}, EIGER \citep{Daichi:2023aa} and GLASS-JWST ERS \citep{Treu:2022aa} have utilized the WFSS mode to study a wide range of science topics including star-formation \citep{Matharu:2024aa, Lu:2025aa}, galaxy environment \citep{Witstok:2024aa}, and metallicity \citep{Wang:2022aa, He:2024aa, He:2026aa}. 
 
 Pure-parallel observing mode\footnote{In this mode, a secondary instrument operates simultaneously with the primary instrument, observing a different part of the telescope's focal plane.} offers a powerful alternative for performing large slitless spectroscopy surveys, as opposed to the targeted programs listed above. Such parallel observing strategies have also been exploited by the \hst to great success \citep{Pasquali:2003aa, Pirzkal:2004aa, Atek:2010aa, Brammer:2012eu, Pirzkal:2013aa, Treu:2015aa, Calvi:2016aa, Pirzkal:2017aa, Henry:2021aa, Wang:2022aa, Battisti:2024aa}. Coupled with \jwst's unique sensitivity and large collecting area, the pure-parallel observing strategy can provide near-infrared data from many areas of the sky.
 
Parallel Application of Slitless Spectroscopy for the Analysis of Galaxy Evolution \citep[PASSAGE;][]{Malkan:2025aa} was the first pure-parallel program approved for NIRISS/WFSS, with more programs approved in the subsequent observing cycles (e.g., PID 3383 with NIRISS/WFSS in Cycle 2, PIDs 5398, 6434 with NIRCam/WFSS in Cycle 3). PASSAGE yielded spatially resolved spectra and imaging of galaxies in each of its $\ang{;2.2;}\times\ang{;2.2;}$ fields of view, at a pixel scale of $\ang{;;0.066}$/pixel. As such, several extragalactic studies with a large sample are enabled by PASSAGE, including discovering Lyman-$\alpha$ emitters \citep{Runnholm:2025aa} at very high redshifts, extending the mass-metallicity relation to low galaxy masses (Nedkova et al., in prep), dust-extinction (Colbert et al., in prep) and spatially resolved properties of clumps (Hasan et al., in prep).

The goal of this work is to demonstrate the utility of NIRISS/WFSS data in studying spatially resolved abundances of individual galaxies at $z\sim2-3$, particularly at the low-mass end (8.0 $<$ \logM $<$ 9.5). In particular, we chose only those galaxies -- from PASSAGE and GLASS-JWST ERS -- with the best quality data (see \cref{sec:sample}). We employ state-of-the-art data reduction and modeling techniques to study spatially resolved metallicities. Our objective is to demonstrate the best scientific value achievable with currently available models and NIRISS/WFSS data, as well as to uncover its limitations, in terms of understanding galaxy evolution through the metallicity and star-formation maps.

This paper is organized as follows. In \cref{sec:data} and \cref{sec:sample} we describe the data and sample selection for both PASSAGE and GLASS-JWST. We outline the analysis methods we adopted, including the various metallicity diagnostics used, in \cref{sec:methods}, followed by laying out the results in \cref{sec:results}. We discuss the implications and caveats of our work in \cref{sec:disc} and summarize our conclusions in \cref{sec:sum}. A flat $\Lambda$CDM cosmology \citep{Planck:2014aa} was used, with $1 - \Omega_\lambda = \Omega_m = 0.285$, $\Omega_b = 0.0461$, and $h=0.695$. We used a solar oxygen abundance of \logOH = 8.76, which is implicit in the photoionization models used by NebulaBayes \citep{Thomas:2018aa} and was derived from the `local galactic concordance' reference values from \citet{Nicholls:2017aa}. 

\section{The data}
\label{sec:data}
We derived our sample from two surveys performed with NIRISS in WFSS mode -- PASSAGE and GLASS-JWST. In this section we describe the surveys and their data reduction procedures.

\subsection{PASSAGE survey}
Our study of the spatially resolved interstellar medium (ISM) of individual galaxies necessitates deep, multi-filter and multi-orientation WFSS observations. The durations of pure-parallel observations are fixed by the primary observation slots. This can sometimes severely limit (a) the exposure duration per grism orientation per filter, (b) the number of grism orientations used, and/or (c) the number of filters used -- progressively reducing the diagnostic power of the data for spatially resolved studies. One way to mitigate shallow observations is by stacking the spectra of multiple galaxies. However, upon stacking one trades off useful information at the level of the individual galaxies in exchange for a statistically significant sample of spatially-averaged properties, which we wish to avoid. Availability of multiple grism orientations helps associate slitless spectra to the corresponding spatial locations robustly, by minimizing confusion due to overlapping spectra.  Thus obtaining grism observations with spectra running in two orthogonal directions (along rows or along columns) helps provide robust 2-D emission line maps \citep{Pirzkal:2024aa}. Additionally, spectra from multiple filters, naturally increase the wavelength coverage, and consequently, the chances of detecting nebular emission lines.

PASSAGE is a \jwst Cycle 1 pure-parallel survey using NIRISS/WFSS \cite[PID: 1571, PI: M. Malkan;][]{Malkan:2025aa}. It observed 63 independent fields, each with a $\ang{;;133}\times\ang{;;133}$ field of view. Most observing blocks for PASSAGE were short ($\lesssim2$ hours), thereby allowing spectroscopic observation with only one filter and a single grism orientation for most fields. For our purposes we needed the WFSS data to cover grisms with both orientations `GR150R' (Rows) and `GR150C' (Columns) in all 3 NIRISS filters -- F115W, F150W and F200W. We also required overlap with COSMOS-Web \jwst/NIRCam footprint, to leverage the COSMOS-Web photometry \citep{Shuntov:2025aa} for stellar mass estimates. Just one PASSAGE field, Par028, satisfied all of the above criteria, so we limited our analysis to this field. Since the objective of this work is to demonstrate spatially resolved science with the best possible data from PASSAGE rather than studying a representative galaxy population, the field-to-field variation introduced by using just one PASSAGE field is of little consequence.

Par028 was observed on 19$^{\rm th}$ and 20$^{\rm th}$ May 2023. The exposure times of Par028 for various combinations of filters and grism or direct imaging modes are given in \cref{tab:par28_exptime}. The total exposure for slitless spectroscopy data on this field is $\sim34$ hours.

\begin{table}
    \centering
    \begin{tabular}{c|c|c|c|c}
    \toprule
    Filters & G150R & G150C & Direct & Total \\
    \midrule
    F115W & $42517$s & $34014$s & $6442$s & $23.0$ hr \\
    F150W & $17007$s & $8503$s & $5154$s & $8.5$ hr \\
    F200W & $12884$s & $6442$s & $5154$s & $6.8$ hr \\
    \midrule
    Total & $20.1$ hr & $13.6$ hr & $4.7$ hr & $38.3$ hr \\
    \bottomrule
    \end{tabular}
    \caption{List of exposure times for various grism orientations and direct images for the Par028 field in PASSAGE.}
    \label{tab:par28_exptime}
\end{table}

\subsection{PASSAGE data reduction}
\label{sec:passage_data_red}

We refer the reader to \citet{Malkan:2025aa} for a description of PASSAGE data reductions. Briefly, the Grism redshift \& line analysis software for space-based slitless spectroscopy \citep[\grizli,][]{Brammer:2023aa} was used for the reduction and spectral extraction, with several customizations tailored for working with PASSAGE data, combined with the grism configuration files provided by the NGDEEP calibration \citep{Pirzkal:2024aa}. Identification of sources was done using ``Source Extraction and Photometry" \citep[\textsc{sep}\footnote{\url{https://sep.readthedocs.io}}; ][]{Barbary:2016aa,Bertin:1996fk}. Then the spectra associated with each detected source were modeled simultaneously for each individual grism exposure. In cases of overlapping spectra, the pixels containing model spectra of the `contaminating object' were down-weighted during the process of 2D grism extraction. \grizli outputs 2D segmentation maps as well as 2D spatially resolved emission line maps and associated weight maps for all lines available in the \grizli templates, for each detected galaxy in the field. Additionally, \grizli provides the best-fit redshift along with integrated emission line fluxes. The integrated fluxes are based on summing the best-fit 2D \grizli templates within the segmentation map of a given galaxy.

\subsection{GLASS-JWST survey}
\label{sec:glass}
Abell 2744 is a well-observed massive galaxy cluster at $z\sim 0.3$, known for its strong-lensing features. To leverage the significant ancillary dataset available for this field, the GLASS-JWST Early Release Science (ERS) program \citep[PID: JWST-ERS 1324, PI: T. Treu, hereafter GLASS for conciseness;][]{Treu:2022aa} obtained imaging of Abell-2744 in multiple NIRCam bands \citep{Merlin:2022aa, Paris:2023aa}, as well as high-resolution spectra with NIRSpec \citep{Mascia:2024aa, Li:2025aa} and low-resolution spectra with NIRISS in the WFSS mode \citep{Boyett:2022aa, Roberts-Borsani:2022aa, Watson:2025aa}. As such, the GLASS-NIRISS data are similar to those from the PASSAGE survey. Their observations comprise of $\sim3$ hours of slitless spectroscopy in each grism orientation for F115W and F150W and $\sim1.5$ hours per orientation for F200W, amounting to $\sim15$ hours of total exposure time, which is comparable to that of the Par028 PASSAGE field.

\subsection{GLASS data reduction}
\label{sec:glass_data_red}

\citet{Watson:2025aa} provide a detailed description of the GLASS NIRISS observations and data reduction, but we highlight the key aspects relevant for our work here. Similar to PASSAGE, the GLASS data were reduced using the \grizli software, albeit with certain customizations to appropriately account for the intra-cluster light owing to Abell-2744 being a cluster field. The grism trace configuration of \cite{Pirzkal:2024aa} was judged to be insufficiently accurate in the 0th order, which is crucial for a crowded field such as Abell 2744\footnote{This does not pose a significant challenge for non-crowded fields, so using these configurations for PASSAGE data was deemed acceptable.}. Therefore the *221215.conf grism configuration \citep{Matharu:2022aa} was used instead, along with the operational context of “jwst\_1173.pmap”. Source detection was performed on mosaics drizzled to a scale of \ang{;;0.03}/pixel. The GLASS data reduction yielded the same end products as that of PASSAGE.

\section{The Sample}
\label{sec:sample}
In this section, we describe our criteria to select the sample of galaxies for this work in \cref{sec:sample_sel}. We also give an overview of the global properties of the selected sample in \cref{sec:global} and present our novel approach for distinguishing star-forming galaxies from AGN in \cref{sec:ohno}.

\subsection{Sample selection}
\label{sec:sample_sel}

\begin{figure*}
    \centering
    \begin{subfigure}[t]{0.49\linewidth}
        \includegraphics[width=\linewidth, trim=1in 0.1in 0.2in 0.1in]{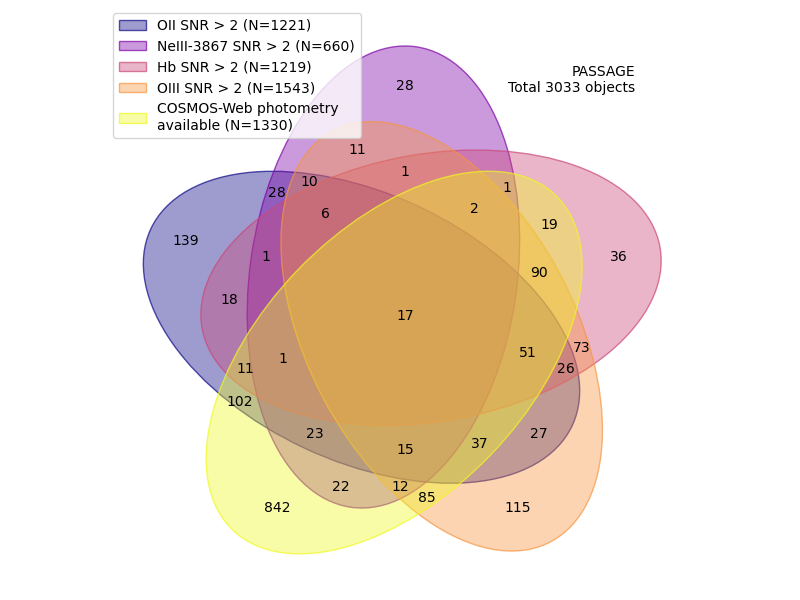}
        \caption{}
        \label{fig:venn_passage}
    \end{subfigure}
    \hfill
    \begin{subfigure}[t]{0.49\linewidth}
        \includegraphics[width=\linewidth, trim=1in 0.1in 0.2in 0.1in]{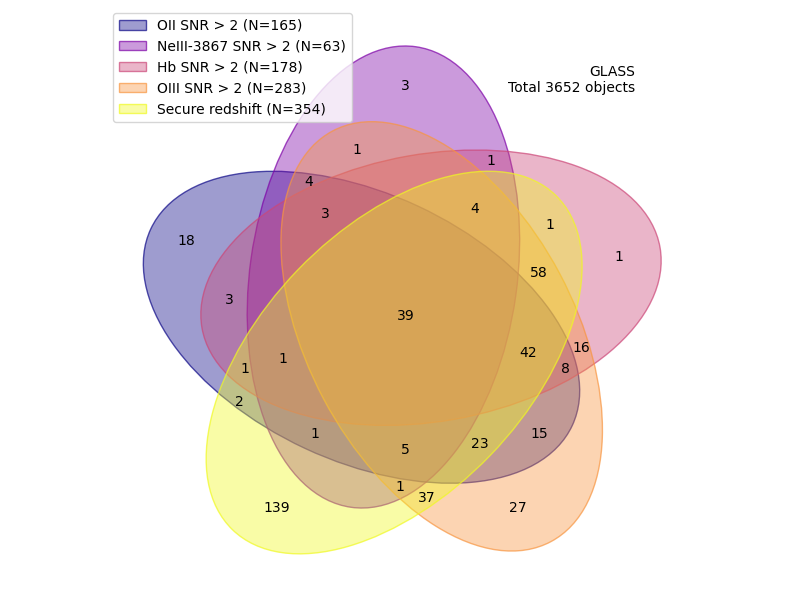}
        \caption{}
        \label{fig:venn_glass}
    \end{subfigure}

    \caption{Venn diagram showing the first step of sample selection within (a) the Par028 field of PASSAGE and (b) the NIRISS field in GLASS. We chose galaxies that have signal-to-noise ratio (S/N) $>2$ in the \grizli-reported integrated line fluxes for each of \oii, \neiii, \hb and \oiii lines. We imposed an additional stellar mass availability requirement for PASSAGE, and a secure redshift criterion for GLASS (see \cref{sec:sample_sel}). The Venn diagram demonstrates how many galaxies were selected based on the above criteria, as well as how many were missed based on individual criteria. The number of galaxies picked by each criteria individually is noted in the legend. The total number in the sectors of the Venn diagram may not add up to the total number of objects quoted on the top right of each panel, because there can be objects that do not satisfy any of the criteria and are therefore not counted in any sector. Overall, this figure depicts the first step of our sample selection.}
    \label{fig:venn}
\end{figure*}

Our goal is to select the best candidates for this spatially resolved study from PASSAGE and GLASS. This involves first selecting the galaxies with relevant emission lines, and then discarding galaxies with potential active galactic nuclei (AGN). We describe the first step here, and the second step, which relies on the \neiiilam line, in \cref{sec:ohno}. For the spatially resolved spectroscopy, we limited our analysis to the \oiilam, \oiiilam, and \hbeta emission lines, because these strong lines are sufficient to determine nebular metallicity. In the faint outer parts of some galaxies the \siilam line appears to have been artificially enhanced by morphological broadening  of the bluer and brighter \ha line (Watson et al. in prep). We therefore did not use the \sii line in this paper. Moreover, we did not require the availability of the \halpha line for measuring metallicity, which allows us to probe redshifts $z > 2.4$. However, wherever the \halpha+\nii complex was available, we used it to compute the star-formation rate (SFR), after correcting for reddening and for the blended \nii flux using the measured metallicity (see \cref{sec:sfr}). In the absence of \halpha, we used de-reddened \oii for measuring SFR (\cref{sec:sfr}). 

We followed a 2-step process for selecting our sample, for both PASSAGE and GLASS -- an automatic filtering, shown in \cref{fig:venn}, followed by visual inspection. In the first step, we selected all galaxies with S/N $>2$ in the integrated fluxes for each of the four emission lines -- \oii, \oiii, \hbeta and \neiii -- as reported by \grizli. We imposed a fifth criterion for Par028 -- availability of COSMOS-Web photometry based on cross-matching between COSMOS-Web and PASSAGE galaxies. We include the COSMOS-Web photometry in estimating stellar masses (see \cref{sec:mstar}) of Par028 galaxies. The same criterion would be redundant for GLASS, since photometry for all  GLASS galaxies is already available via the MegaScience \citep{Suess:2024aa} and UNCOVER surveys \citep{Bezanson:2024aa}. Instead, for GLASS we imposed a different fifth criterion -- availability of secure redshifts. The redshift flag was a product of rigorous visual inspection by the GLASS team (see \citealt{Watson:2025aa} for details). We expect the redshifts of our selected PASSAGE galaxies to be secure, given the multiple strong emission line detections. This is indeed demonstrated in the GLASS Venn diagram (\cref{fig:venn_glass}): out of all the galaxies that have all four lines detected, 39 have secure redshifts while only three do not. Generally, we have fewer galaxies in different bins of the Venn diagram in GLASS compared to PASSAGE, yet we have more GLASS galaxies satisfying our detection criteria. This is likely because: (a) the GLASS spectroscopic observations are shallower than PASSAGE, leading to fewer galaxies with detected lines, and (b) the GLASS galaxies with any of the above lines detected are more likely to have all the above lines detected than PASSAGE, due to a galaxy over-density around $z\sim2$ \citep{Watson:2025aa} that serendipitously puts all the relevant lines within NIRISS wavelength coverage.

\begin{figure}
    \centering
    \includegraphics[width=1\linewidth, trim=1in 0in 2in 0in]{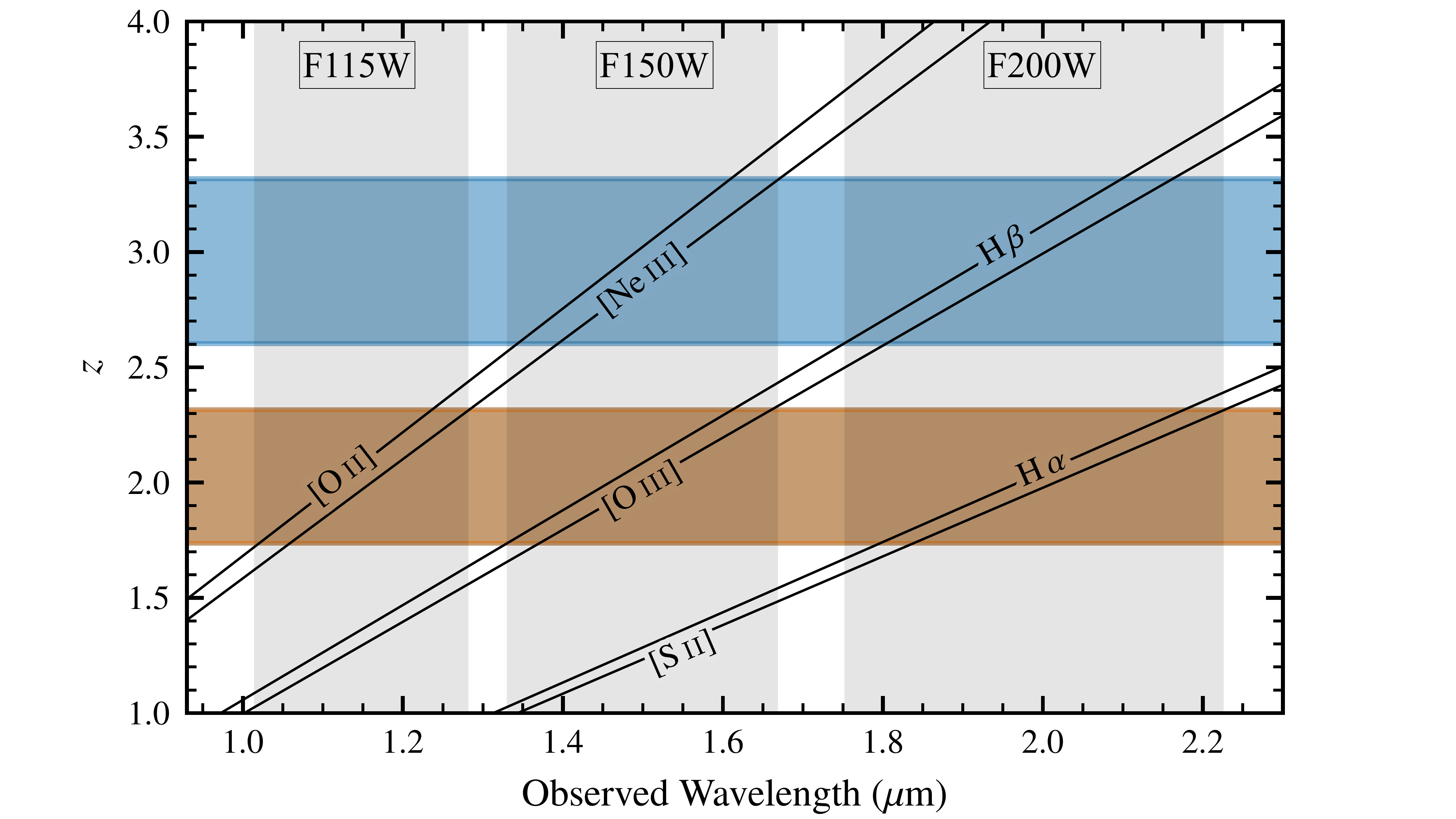}
    \caption{Illustration of the redshift ranges NIRISS can cover while targeting the lines \oii, \neiii, \oiii, \hb, \ha and \sii. Filter coverage (within 50\% of transmission curve) for F115W, F150W and F200W are shown as gray vertical bands from left to right. The brown horizontal band indicates the redshift range where all of the above lines are available to NIRISS, and the blue band indicates the additional redshift-space that becomes accessible upon dropping \ha and \sii.}
    \label{fig:lines_availability}
\end{figure}

The above selection yielded 17 objects in Par028, and 39 in GLASS, as shown in \cref{fig:venn}. \cref{fig:lines_availability} shows the redshift regime in which the relevant emission lines are accessible to NIRISS, which applies to both PASSAGE and GLASS. Considering the gaps and sensitivity variation of each NIRISS filter, our required four lines would, by definition, be detected in adjacent filters, with two in each filter. When \oii and \neiii are captured in F115W and \hbeta and \oiii appear in F150W, we can access a redshift range 1.7 < $z$ < 2.3, denoted by the brown horizontal band in \cref{fig:lines_availability}. In this case, \halpha would be accessible in the F200W filter. Alternatively, if our required four lines happen to lie in the F150W and F200W filters, our redshift access gets extended to 2.6 < $z$ < 3.3, shown by the blue band, albeit missing \halpha. Therefore, not relying on \halpha (or \sii) for this study, effectively extended our access to a higher redshift range.

In the second step, we visually inspected all candidates and discarded those with significant contamination, unreliable line fits (to exclude cases where the \grizli-reported SNR was untrustworthy), and cases where any emission line of interest happened to be too close to the filter edges (i.e. sharply falling sensitivity). While an exact count of galaxies falling under each of the above categories is unnecessary, it is worth noting that the majority of the discards were approximately evenly split between contamination and unreliable line fits, with a minor contribution from lines close to filter edges. Overall, 13 Par028 galaxies and 25 GLASS galaxies were discarded. This left us with four objects in Par028 and 14 objects in GLASS, pending further checks. 

We performed an additional third step for GLASS to avoid strongly lensed systems. Obtaining accurate source-frame 2D metallicity maps of strongly lensed systems would be challenging \citep[see ][He et al. in prep]{Wang:2022aa, Metha:2024aa} and would require employing lens reconstruction techniques, which is beyond the scope of this work. Therefore we discarded multiply-imaged systems and any object with a magnification $\mu>2$ \citep{Bergamini:2023aa}. This yielded four minimally-distorted galaxies in the GLASS field, all of which are $\gtrsim300$ kpc away from the cluster center in projection. We do not correct for magnification for these galaxies. Given that our goal is to essentially correlate emission line maps, and all maps are stretched the same way, that stretching would have minimal impact on said correlation. Essentially our results for the GLASS galaxies can be thought of as image-plane correlations.

In summary, this selection yields eight galaxies -- four from Par028 (PASSAGE) and four from Abell 2744 (GLASS). Given our end goal of measuring metallicity, and the fact that most metallicity diagnostics assume ionization by star-formation, we ensured that our sample does not include any AGN-hosts that might skew our metallicity measurements. Therefore, the next step was to discard potential AGN-hosts, as described in \cref{sec:ohno}.

\subsection{Spatially integrated properties}
\label{sec:global}
To distinguish AGN-hosts from star-forming (SF) galaxies we need emission line flux ratios. But first, we need to define the center and size of the galaxies to determine the extent within which to measure the fluxes, which we describe below.

\subsubsection{Galaxy size and center}
\label{sec:re}

We limited our analysis to the area within the segmentation map of each galaxy, which was derived by \textsc{sep} during the data reduction. Even then, the outskirts of galaxies usually displayed very low S/N. To exclude regions of low signal, we further limited all our analysis to a square of size $\pm$ 5  x 5 $\times$ the effective radius (R$_e$) of the galaxy. 

We assigned the brightest pixel in the smoothed (with a $5 \times 5$ pixel boxcar kernel) direct F150W image as the galaxy center, which we used throughout the paper. Comparison with the segmentation map geometry confirmed that this is a good estimate of the center. To measure R$_e$, we first de-convolved the original direct F150W image using a Richardson-Lucy algorithm \citep{Richardson:1972aa, Lucy:1974aa} in Python's \texttt{skimage} module \citep{scikit:2014aa}, in order to avoid over-estimation of R$_e$ due to PSF-smearing\footnote{Not having corrected for magnification can impact our R$_e$ measurements of the GLASS galaxies \citep{Miller:2025aa}. However, our choice of low-magnification galaxies helped keep this impact small.}. Then we assigned the half-light radius of the deconvolved image as the R$_e$. For our sample we typically find R$_e$ $\sim$\ang{;;0.1}-\ang{;;0.3} ($\sim$ 1-2\,kpc). Compared to the \ang{;;0.092} FWHM of the NIRISS PSF at F150W, all our galaxies are resolved (\autoref{sec:ap_re}). We pick F150W since it is the middle filter and the differences due to PSF (as well as morphological effects) between it and F115W or F200W is minimal. Choosing the F150W filter is also beneficial since this filter has the most similar exposure times in Par028 (5154s) vs GLASS (5669s) which minimizes variation between the two fields.

\subsubsection{Choice of integrated fluxes}
\label{sec:integrated}

Although our primary goal is to study spatially resolved properties, we measured the global values of these properties as well, and put them in the context of the local relations. These global properties (e.g., global line ratios, metallicity, SFR) were derived from integrated flux measurements. Therefore it is important to define what we used as the `integrated line flux' throughout this paper. 

There were two choices -- the integrated flux as reported by \grizli, and the one obtained by summing the 2D emission line maps. Theoretically, these should be identical. While these are not identical in practice owing to the complexities in the extraction of emission maps in \grizli ($\lesssim$0.2 dex scatter), there is no systematic difference between the two (see \cref{sec:ap_int} for a comparison). We chose to use the summed flux of the 2D line maps (within a box size of 5\,R$_e$) as the integrated line flux for each galaxy, because this approach is directly related to the resolved line maps and therefore facilitates the comparisons between spatially resolved and integrated quantities presented later in the paper. 

We corrected all emission line fluxes for dust reddening using the \citet{Cardelli:1989} dust law and the \halpha/\hbeta ratio, wherever \halpha was available. For the de-reddening process, we used a fixed ratio \halpha/(\nii + \halpha) = 0.823 \citep{James:2005aa} to obtain \halpha flux from the \nii + \halpha complex. For the galaxy at $z\sim3$ with no \halpha available we did not perform any dust-correction. Given the typically dust-poor and highly star-forming nature of our sample, by selection, the impact of lack of de-reddening for this galaxy would be small. Indeed, we obtain a strict upper limit on the dust content from SED fitting of this galaxy is $E(B-V)_{\rm stars} = 0.0005 \pm 0.0161$ (using the \citealt{Cardelli:1989} extinction law), corresponding to a nebular $E(B-V)_{\rm neb} = 0.001 \pm 0.037$\footnote{$E(B-V)_{\rm neb} = \frac{E(B-V)_{\rm stars}}{0.44}$, where the factor 0.44 is from \citet{Calzetti:2000vn}}, which is consistent with being dust-free. Lastly, we obtained the integrated as well as spatially resolved \oiiilam flux from the \oiiiablam doublet by assuming a 3:1 ratio between \oiiilam and \ensuremath{\left[\ion{O}{iii}\right]\,{\lambda 4959}}.

\subsubsection{Stellar masses}
\label{sec:mstar}

We measured stellar masses of all galaxies in a consistent manner. First, we cross-matched each Par028 galaxy with the COSMOS-Web catalog \citep{Casey:2023aa} and each GLASS galaxy with UNCOVER-DR3/MegaScience catalog \citep{Suess:2024aa}\footnote{While both the \citet{Merlin:2022aa} and \citet{Paris:2023aa} catalogs contain photometry for the GLASS galaxies, the MegaScience catalog is the only release at present to include all JWST medium band filters.}. The closest neighbor in the external catalog within a \ang{;;0.1} radius was considered a match. This process yielded a median separation of \ang{;;0.04} between pairs of matched objects. For each matched object, we extracted the photometry in all available \hst and \jwst bands, within a \ang{;;1} aperture, from the external catalogs. This led to six bands (ACS F814W, NIRCam F115W, F150W, F277W, and F444W, and MIRI F770W) for the Par028 objects from COSMOS-Web photometry catalog \citep{Shuntov:2025aa} and 26 bands (ACS F435W, F606W, and F814W, WFC3 F105W, F125W, F140W, and F160W, NIRCam F070W, F090W, F115W, F140M, F150W, F182M, F200W, F210M, F250M, F277W, F300M, F335M, F356W, F360M, F410M, F430M, F444W, F460M, and F480M) for the GLASS objects from the UNCOVER catalog. To this we added the photometry within \ang{;;1} aperture in the three NIRISS bands (F115W, F150W, F200W) from PASSAGE and GLASS observations, leading to nine and 29 bands for the two samples respectively.

We used \textsc{BAGPIPES} \citep{Carnall:2018aa} with \citet{Bruzual:2003lr} stellar population synthesis (SPS) models and a \citet{Kroupa:2002aa} initial mass function (IMF), to fit a spectral energy distribution (SED) to the above photometry. We assumed a continuity star-formation history \citep{Leja:2019aa} with a Student's t-width $\sigma = 0.5$, allowing for 5 age bins with bin edges spaced in geometric progression from 30\,Myr to the current age of the galaxy. The total stellar mass formed was allowed to vary between 10$^6$ and 10$^{11}$ \Msun, and metallicity was allowed between 0 and 3\,$Z_{\odot}$. We used the \citet{Cardelli:1989} dust attenuation curve with $\eta=2$ \citep{Buat:2018aa} and $0 < A_V < 2$, where $\eta$ is the multiplicative factor on $A_V$ for to account for differential attenuation between young and old stars. The nebular emission component of the models were also utilized for the SED fitting, allowing the ionization parameter to vary within $-3.5 < \log{U} < -2.0 $. Here, $\log{U}$ is the dimensionless representation of ionization parameter ($U = q/c$, where $c$ is the speed of light), and the above limits correspond to $7.0 < \log{q\,({\rm cm/s})} < 8.5 $.

\subsection{Identifying star-forming galaxies: the OHNO diagram}
\label{sec:ohno}
The commonly used diagnostics for demarcating SF galaxies vs AGN-hosts are the \citet[][hereafter BPT]{Baldwin:1981rr} or \citet[][hereafter V087]{Veilleux:1987aa} diagrams, which employ the parameter space of the \oiii/\hbeta ratio vs \nii/\halpha or \sii/\halpha line ratios, respectively. However, the lack of resolved \nii and \ha lines in NIRISS spectra, and the unreliability of the \sii lines render the traditional BPT and V087 diagrams ill-suited for this analysis.

\begin{figure*}
    \centering
    \includegraphics[width=0.33\linewidth]{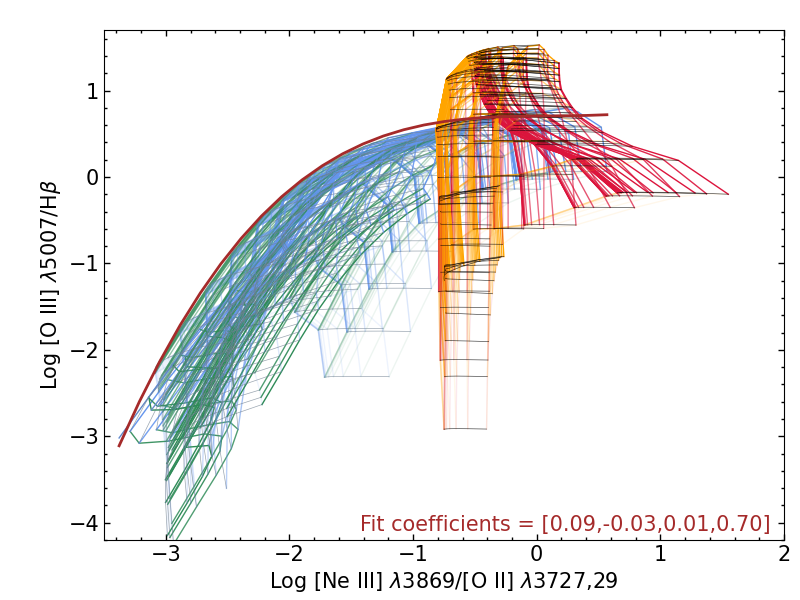}
    \includegraphics[width=0.32\linewidth]{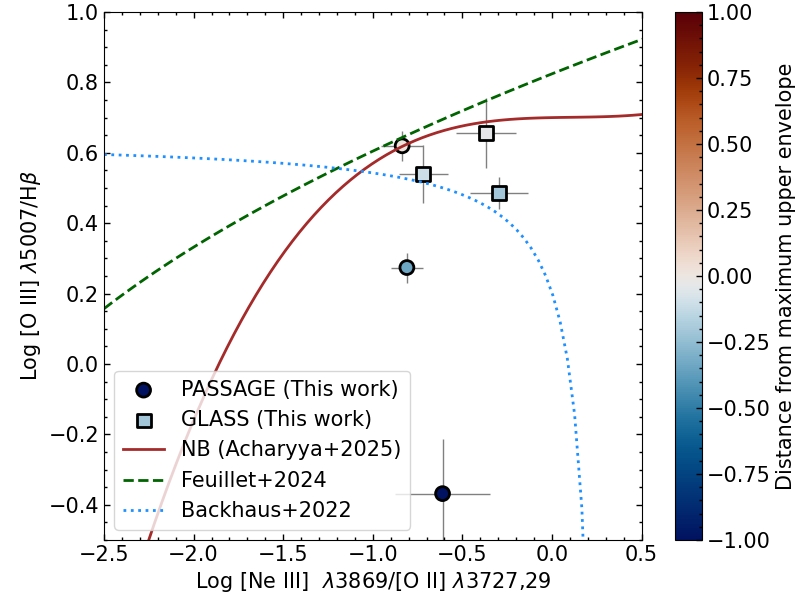}
    \includegraphics[width=0.32\linewidth]{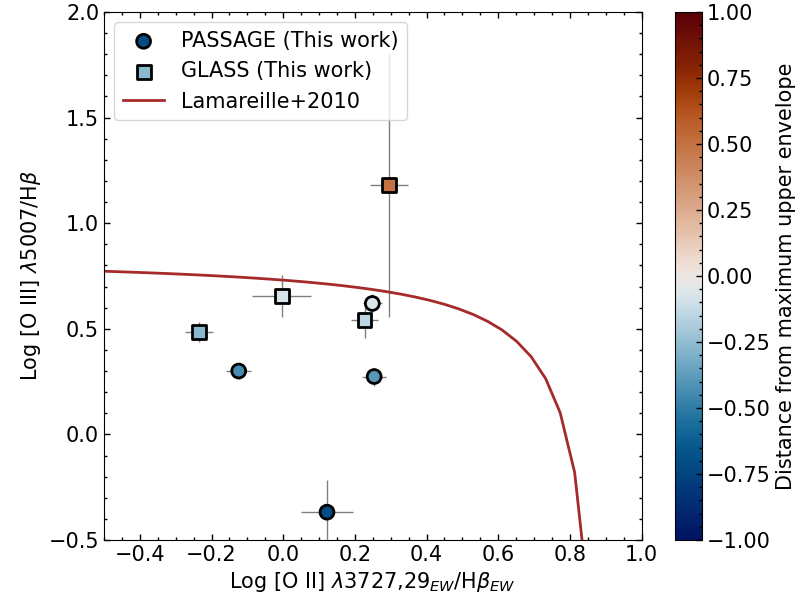}
    \caption{Left: \ohno diagram based on the MAPPINGS v5.1 \hii region model grids \citep{Thomas:2018aa}. Each green and blue line connects grid points with same values of metallicity (\logOH) and ionization parameter (\logU), respectively, with the line opacity scaling linearly with the values of \logOH and \logU. We deliberately did not annotate the different lines with \logOH and \logU values to avoid over-crowding, because our goal is simply to demonstrate the existence of an upper envelope of this grid. The solid red line denotes the fourth order polynomial fit to this upper envelope. The polynomial coefficients are in the bottom right. \edit{Additionally, we show the AGN grid from \citet{Thomas:2018aa} in the same parameter space, but with orange and crimson colors instead of green and blue, respectively (see \cref{sec:ohno} for details)}. Middle: Illustration of application of the AGN demarcation line on the integrated flux ratios of PASSAGE (circle) and GLASS (square) galaxies. The red line is the same as that from the left panel. The other lines denote existing demarcations from the literature. We deliberately chose different axes limits compared to the left panel, to zoom in on the parameter space relevant to the observed data. Each data point is color-coded by the shortest distance from our AGN-SF demarcation line, with bluer colors for data lying below the line, and redder colors for those above the line. Two of the eight galaxies in our sample had negative \neiii flux when summed within a $\pm$2.5 R$_e$ box and therefore could not be plotted here. Given such low level of \neiii we assume those galaxies are star-forming (SF). Overall, this figure depicts our new method of demarcating AGN vs SF galaxies using the \neiii/\oii ratio. \newedit{Right: Verification using the blue-BPT diagram, i.e., \oiii/\hbeta vs \oii/\hbeta ratio plot \citep{Lamareille:2010aa} that our sample is indeed dominated by star-formation. While one GLASS galaxy has a median value of the \oiii/\hbeta ratio that puts it in the AGN-dominated part of the parameter space, it is nevertheless consistent with being SF-dominated within 1$\sigma$ uncertainties. We therefore retain all eight galaxies in our sample.}}
    \label{fig:nb_fit}
\end{figure*}

We hence employ the \ohno diagram -- \oiii/\hb vs \neiii/\oii parameter space -- to identify the galaxies whose emission is dominated by processes other than SF\footnote{Although \neiii is blended with \heilam at NIRISS resolution it does not impact the SF-AGN demarcation for our sample (\cref{sec:ap_hei}).}. This diagnostic was first used by \citet{Zeimann:2015aa} to separate AGN-dominated versus SF-dominated galaxies, and subsequently by other studies including \citet[][hereafter \citetalias{Feuillet:2024aa}]{Feuillet:2024aa}, \citet[][hereafter \citetalias{Backhaus:2022aa}]{Backhaus:2022aa} and \citet{Malkan:2026aa}. However, the demarcation lines provided by \citetalias{Backhaus:2022aa} and \citetalias{Feuillet:2024aa} were empirically derived rather than theoretically modeled. We therefore propose a new AGN demarcation line in the \ohno diagram that is physically motivated.

\cref{fig:nb_fit} depicts our method. The left panel shows the grid of MAPPINGS v5.1 \hii region photoionization models \citep{Thomas:2018aa, Kewley:2019ab}, encompassing 12 values of metallicity within 7.06 $\leq$ \logOH $\leq$ 9.3, 12 values of gas pressure 4.2 $\leq$ \lpok $\leq$ 8.6 and 11 values of ionization parameter -4.2 $\leq$ \logU $\leq$ -0.2. The ionizing spectra for these models were computed using \texttt{slug2} \citep{Krumholz:2015aa} stellar population synthesis code with the following settings: a single stellar population after 10 Myr of continuous star-formation, Chabrier IMF, and Padova stellar tracks. The MAPPINGS models indicate that for a given \neiii/\oii ratio, there is a maximum limit to which \oiii/\hb ratio can be reproduced by \hii regions. Therefore, the AGN-SF test mainly depends on how high the \oiii/\hb ratio is--if it exceeds the maximum envelope of this grid of models, it must have originated from regions dominated by non-SF ionizing sources. While there is some debate about the redshift dependence of the \oiii/\hbeta ratio \citep[e.g.,][]{Garg:2022aa}, the locus of of the $z\sim2$ galaxies in the Keck Baryonic Structure Survey \citep[KBSS;][]{Strom:2017aa} lies well within the \oiii/\hbeta values of our model grid. 

\edit{The left panel of \cref{fig:nb_fit} also shows an AGN grid from \citet{Thomas:2018aa} (orange and crimson lines). We refer the reader to \citet{Thomas:2018aa} for details of these AGN grids, but briefly, these grids were computed using an \texttt{OXAF} ionizing spectrum \citep{Thomas:2016aa}, \citet{Jenkins:2009aa} dust depletion factors, no dust destruction, same range of \logOH, \logU, \lpok as above, and $-2.0 \leq \log E_{\rm peak} \leq -0.75$, where $\log E_{\rm peak}$ is the energy of the peak of the accretion disk emission. To avoid overcrowding, we show only the grid corresponding to $\log E_{\rm peak}=-0.75$ because it occupies the most distinct  \ohno parameter space compared to the \hii region grid. Our objective is not to provide a method to distinguish pure-AGNs from purely SF galaxies (which is challenging, given the overlap of the \hii region and AGN models) but rather to provide a diagnostic to eliminate sources with ionization inconsistent with pure SF, i.e., above \hii region grids in this figure.}

\edit{We can  use the \hii region grid to quantify the AGN-SF demarcation applicable in cases where \nii and \halpha are unavailable (e.g., in low resolution grism spectra). Strictly speaking, the observed \ohno ratios of some of our galaxies could be consistent with so-called ``composite" Seyfert + SF emission line spectra \citep{Malkan:2026aa} but we allow them to remain in our sample on the assumption that star formation is their predominant photoionization process.} The thick red line in the left panel of \cref{fig:nb_fit} is a fourth-order polynomial fit to the upper envelope of the MAPPINGS \hii region models. The fit is given by a 1D polynomial of the form

\begin{equation}
    y = 0.09x^3 - 0.03x^2 + 0.01x + 0.7
\end{equation}

\noindent where $y= \log{(\oiii/\hb)}$ and $x = \log{(\neiii/\oii)}$. We reproduce this fitted line in red, along with the \citetalias{Backhaus:2022aa} and \citetalias{Feuillet:2024aa} lines, on the right panel of \cref{fig:nb_fit}, and overlay the integrated observed line ratios for our sample. Owing to the insufficient S/N in the \neiii emission line maps in our data (even upon applying various spatial binning techniques) we use this diagnostic only with spatially integrated flux ratios. \edit{Strictly speaking, the emission lines of some (mainly early-type) galaxies can be produced not by star-formation but by so called LINERs, which typically have \oiii/\hbeta ratios of 3 or less \citep{Malkan:2026aa}. However they also have \neiii/\oii ratios below 0.07, so it is unlikely that any galaxy in our sample galaxies could be a possible LINER.}

\newedit{To verify this further, we employed the `blue-BPT' diagram, which is used to separate AGN-hosts from SF galaxies in the \oiii/\hbeta vs \oii/\hbeta parameter space. \citet{Lamareille:2010aa} noted that while the wide separation in the \oii and \hbeta wavelengths could lead to dust-reddening having an impact on the \oii/\hbeta flux ratio, the ratio of their equivalent widths (EWs) alleviates this impact somewhat. We therefore used the EWs for this line ratio, as shown in the right panel of \autoref{fig:nb_fit}. All our galaxies are consistent with being SF-dominated within 1$\sigma$ uncertainties in the `blue-BPT', which agrees with our OHNO classification. While the `blue-BPT' provides a useful way to separate AGN vs SF galaxies, in the event of the \halpha line being unavailable (for high-redshift galaxies, for instance) it is challenging to correct the line ratios for dust reddening without the \halpha/\hbeta Balmer decrement measurement. The OHNO demarcation, on the other hand, employs ratios of lines with short wavelength separations, such that the effect of dust-reddening is negligible. This makes the OHNO classification usable even in the absence of \halpha observations, which can occur, for instance, for 2-filter NIRISS observations in certain redshift ranges (see \autoref{fig:lines_availability}).}

Our OHNO demarcation also agrees with the commonly used mass-excitation (MEx) diagram \cite[e.g.,][see \cref{sec:ap_mex}]{Juneau:2013aa, Henry:2013aa, Henry:2021aa}. However, whilst the latter can only distinguish AGN-hosts and SF galaxies globally, our new demarcation line should enable future studies with higher S/N \neiii maps to differentiate between ionization sources in a spatially resolved manner.
We have confirmed that the same SF/AGN classifications of our 9 galaxies would have been obtained using their integrated \oiii/\hbeta and stellar mass values.

In summary, our final sample consisted of eight SF galaxies, all of them being extended sources, where we can reliably extract spatially resolved ISM properties of individual galaxies, without resorting to stacking or source-plane reconstruction. One of these is at  $z\sim3$ (PASSAGE), with the rest at $z\sim2$.

\subsection{Star-forming main sequence relation}
\label{sec:sfms}

\begin{figure}
    \centering
    \includegraphics[width=1\linewidth]{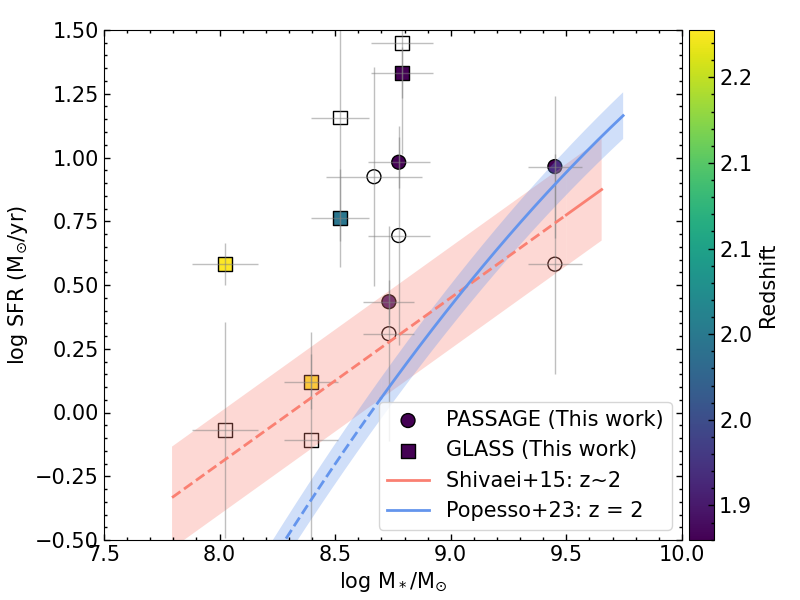}
    \caption{Star-formation main sequence (SFMS) plot depicting the global stellar mass and SFR of the PASSAGE and GLASS galaxies with circles and squares, respectively. The filled symbols denote \ha-based SFR and are color-coded by redshift, whereas the empty symbols denote \oii-based SFR measurements. The \citet{Shivaei:2015aa} and \citet{Popesso:2023aa} lines show the SFMS from existing studies at $z\sim2$, with their extrapolations into lower stellar masses depicted as dashed lines. The \citet{Whitaker:2014aa} calibration, not shown here to avoid crowding, is consistent with the \citet{Popesso:2023aa} line in this stellar-mass regime. Seven of our eight galaxies have \ha available and therefore appear as both filled and empty symbols, which generally agree within their uncertainties. The remaining galaxy is redshifted beyond \halpha coverage with NIRISS and therefore appears only as an empty symbol (i.e. \oii-based SFR). This figure demonstrates that the galaxies in our sample are all very actively star-forming.}
    \label{fig:sfms}
\end{figure}

To place our sample in the context of the general galaxy population at $z\sim2-3$, we plot our galaxies on the SFR-M$_*$ diagram, as shown in \cref{fig:sfms}. Because not all of our galaxies had \halpha available, we measured SFR in two ways: (a) using \halpha, corrected for \nii using metallicity maps, wherever \halpha was available and (b) using \oii for all galaxies. Sect. \ref{sec:sfr} describes both methods in detail. Note that while \oii, in principle, probes similar recent star-formation timescales (i.e., the $\sim$few Myr lifetime of massive OB stars) as \halpha, its dependency on SFR has more scatter than \halpha, due to additional dependency on other factors including metallicity and gas density. The \halpha-based SFRs and \oii-based SFRs are shown by filled and empty points respectively in \cref{fig:sfms}. Generally, the two methods agree within uncertainties. We therefore use the \oii-based SFR for the one galaxy where \halpha is unavailable, for the rest of our analysis.

Comparing with empirical relations from \citet{Shivaei:2015aa} and \citet{Popesso:2023aa}, all of our galaxies are \newedit{either consistent with, or $<$1 dex above,} the SFMS relation at $z\sim2$. Both these relations are based on \halpha-based SFRs, the same as all but one galaxy in our sample. The active star-formation in these galaxies is unsurprising because our selection of S/N$>2$ on four nebular emission lines is expected to bias us towards highly star-forming and less dusty galaxies. We present a complete list the global properties of our sample in \cref{tab:sample}.

\begin{table*}
    \centering
    \caption{List of global galaxy properties of our sample.}

    \begin{threeparttable}

    \begin{tabular}{lrlllllll}
    \toprule
    Field & ID & RA & Dec & $z$ & $\log$ M$_{\star}$/M$_{\odot}$ & SFR (M$_{\odot}$/yr) & $\log$ O/H + 12$_{total}$ & $\nabla Z$ (dex/R$_e$) \\
    \midrule
    Par028 & 300 & $ 150.0893$ & $ 2.4031$ & $ 1.9$ & $ 9.5 \pm  0.1$ & $ 9.21 \pm  5.91$ & $ 8.95 \pm  0.83$ & $-0.53 \pm  0.52$ \\
    Par028 & 1303 & $ 150.0971$ & $ 2.4159$ & $ 1.9$ & $ 8.8 \pm  0.1$ & $ 9.58 \pm  2.20$ & $ 8.42 \pm  0.35$ & $-0.01 \pm  0.28$ \\
    Par028 & 1849 & $ 150.0939$ & $ 2.4220$ & $ 3.1$ & $ 8.7 \pm  0.2$ & $ 8.41 \pm  8.32$\tnote{a} & $ 7.77 \pm  0.15$ & $ 0.07 \pm  0.14$ \\
    Par028 & 2867 & $ 150.0862$ & $ 2.4350$ & $ 2.0$ & $ 8.7 \pm  0.1$ & $ 2.72 \pm  0.54$ & $ 8.60 \pm  0.12$ & $ 0.45 \pm  0.24$ \\
    glass-a2744 & 1721 & $ 3.6061$ & $-30.3935$ & $ 2.2$ & $ 8.4 \pm  0.1$ & $ 1.32 \pm  0.33$ & $ 8.36 \pm  0.15$ & $-0.05 \pm  0.12$ \\
    glass-a2744 & 1983 & $ 3.6133$ & $-30.3911$ & $ 1.9$ & $ 8.8 \pm  0.1$ & $ 21.51 \pm  4.99$ & $ 8.36 \pm  0.47$ & $ 0.26 \pm  0.36$ \\
    glass-a2744 & 1991 & $ 3.6032$ & $-30.3911$ & $ 2.2$ & $ 8.0 \pm  0.1$ & $ 3.83 \pm  0.71$ & $ 7.83 \pm  0.12$ & $ 0.08 \pm  0.20$ \\
    glass-a2744 & 1333 & $ 3.6165$ & $-30.3978$ & $ 2.0$ & $ 8.5 \pm  0.1$ & $ 5.78 \pm  2.57$ & $ 7.89 \pm  0.77$ & $-0.53 \pm  0.33$ \\
    \bottomrule
    \end{tabular}

    \begin{tablenotes}
    \footnotesize
    \item[a] This SFR is derived from \oii luminosity, while the rest are from \halpha (\cref{sec:sfms}). The high uncertainty is a result of propagating the uncertainties in the exponent of \citet{Figueira:2022aa} calibrations.
    \item General note: From left to right the columns are: (a) Observed field, (b) galaxy identifier (derived from PASSAGE and GLASS-NIRISS catalogs), (c, d) on-sly coordinates of the galaxies (e) redshift, as reported by \grizli, (f, g) stellar mass and SFR, as determined in \cref{sec:mstar} and \cref{sec:sfr}, respectively, (h, i) integrated metallicity and metallicity gradient values determined using \texttt{NebulaBayes} (see \cref{sec:zdiag_nb}).
    \end{tablenotes}
    \end{threeparttable}
    \label{tab:sample}
\end{table*}

\section{Spatially resolved quantities}
\label{sec:methods}
In this section we describe our steps and diagnostics used to obtain the spatially resolved metallicity and SFR maps.

\subsection{Spatial binning}
\label{sec:radbin}

Given the generally low S/N per pixel in our data, we chose to radially bin the 2D emission line maps in 5 non-uniformly spaced elliptical bins\footnote{We investigated using Voronoi binning on our data but that led to extremely patchy metallicity maps hindering a reliable radial profile measurement.} -- 0.2, 0.6, 1.2, 1.8 and 2.5 R$_e$ along major-axis radii. The choice of bin size was to ensure optimal SNR in all the resulting annuli, but a different choice (e.g., uniformly spaced) do not impact our science results. The elliptical bins were computed using the corresponding position angle and ellipticity of the galaxy, as reported by \grizli's source-extraction. The flux uncertainties were appropriately propagated during this binning process. We repeated the same binning process using the same set of bins for all the emission line maps of a given galaxy.

\begin{figure*}
    \centering
    \includegraphics[width=0.8\linewidth]{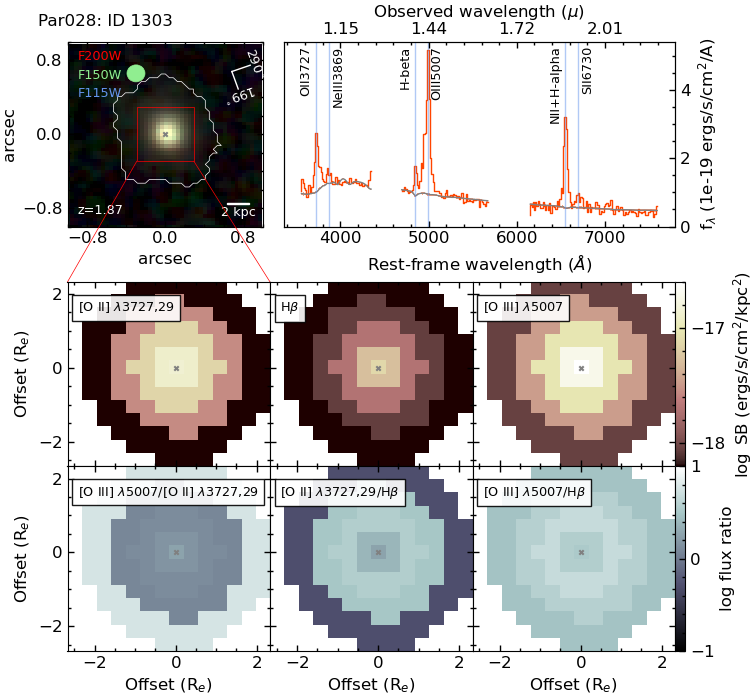}
    \caption{One of our PASSAGE galaxies, as an example of spatially resolved line emission. \textit{Top left:} \ang{;;1}$\times$\ang{;;1} RGB cutout using the direct images in filters F115W, F150W and F200W. The FWHM of the PSF in the F150W filter is denoted as a circle of corresponding color. The white contour denotes the extent of the segmentation map, up to which we limit all our analysis. The position angles for the dispersion directions, corresponding to R and C grism orientations, are denoted in the top right, with the redshift and scale bar at the bottom.  The red box denotes a square of side 5\,R$_e$ around the galaxy center, which we zoom in to for measuring all line maps. \textit{Top right:} The integrated 1D spectra are shown in orange, with corresponding uncertainty as shaded background (it is hard to see due to the small relative uncertainties), and the stellar continuum is shown in gray. The wavelength locations of emission lines of interest are annotated. \textit{Middle row:} Zoomed-in, binned 2D emission line surface brightness (SB) maps for (from left to right) \oiilam, \hb, \oiiilam. \textit{Bottom row:} Emission line ratio maps corresponding to (from left to right) \oiii/\oii, \oii/\hb and \oiii/\hb. Overall, this figure illustrates the result of our spatial binning for an individual galaxy.}
    \label{fig:galaxy}
\end{figure*}

\cref{fig:galaxy} presents one of our galaxies -- ID \#1303 from Par028 -- as an example of the resulting binned emission line surface brightness maps and flux ratio maps. Here we show the RGB image -- generated using drizzled mosaics with filters F200W (R), F150W (G) and F115W (B) -- and the integrated 1D spectrum in the top row, 2D emission line maps and line ratio maps in the bottom row. We use this same galaxy for illustrative examples throughout the paper, wherever needed.

\subsection{Bayesian metallicity inference}
\label{sec:zdiag_nb}

We employed a Bayesian framework to determine metallicity. Several previous studies \citep[e.g.,][]{Jones:2015ab, Wang:2017aa, Henry:2021aa, Wang:2020aa, Wang:2022aa, Venturi:2024aa, Revalski:2024aa} have successfully demonstrated that a Bayesian approach is useful to reduce dependence on any individual strong emission line (SEL) calibration\footnote{We nevertheless compare metallicity values obtained using \citet{Cataldi:2025aa} calibrations in \cref{sec:zdiag_sel}.}.

We employed the Bayesian tool -- \texttt{NebulaBayes} -- developed by \citet{Thomas:2018aa}. This compares observed emission line flux ratios with a grid of existing photoionization models and computes the likelihood for the physical conditions at each grid point to have emitted the set of observed fluxes. It then multiplies the likelihood with the user-defined prior to obtain the posterior distribution for each of the three parameters -- metallicity (\logOH), ISM pressure (\lpok) and ionization parameter (\logU) -- simultaneously. We assumed uniform logarithmic priors; so the posterior and likelihood are identical. This is a standard and reasonable assumption, particularly because we do not have information about any of the three properties outside of the observed emission lines which go into \texttt{NebulaBayes}. Note that uniform priors are not always uninformative; they are simply suited here because we do not have an independent source of information to be able to use informed non-uniform priors. \texttt{NebulaBayes} uses a grid of MAPPINGS v5.1 \hii region models, described in \cref{sec:ohno}\footnote{Although \texttt{NebulaBayes} is capable of accepting any arbitrary grid of \hii region models, we chose to use the default models since those are still the most up-to-date MAPPINGS models}. \texttt{NebulaBayes} is capable of accepting the following emission line fluxes relevant to this study: \oiiiablam, \hb, \oiilam, \neiiilam, \niilam\,and \ha. However, since we did not have separate \nii and \ha fluxes, we provided the corrected \ha flux to \texttt{NebulaBayes} using a constant correction fraction \halpha/(\nii + \halpha) = 0.823 \citep{James:2005aa}.

Once provided with a set of observed emission line fluxes and associated uncertainties, \texttt{NebulaBayes} computed the 2D as well as the marginalized probability distribution functions (PDFs) of metallicity, ionization parameter and ISM pressure simultaneously. For each spatial bin of each galaxy we input all positive line flux measurements into \texttt{NebulaBayes}, along with corresponding uncertainties, prior to dust-correction because \texttt{NebulaBayes} internally accounts for dust-reddening. This results in different sets of emission lines\footnote{Typically a subset of 5-8 lines from \oiilam, \neiiilam, H$\delta$, H$\gamma$, \hbeta, \oiiiablam, $\ion{He}{i}\,{\lambda 5876}$, and $\left[\ion{O}{i}\right]\,{\lambda 6300}$, \halpha were provided to \texttt{NebulaBayes} for each spatial bin.
}for different galaxies, although \oiii, \hb and \oii were almost always included. We adopted the peak of the marginalized posterior PDF of metallicity, as the metallicity and set the corresponding uncertainty to be equal to the mean difference of the peak from the lower and upper bound of 68\% confidence interval. We repeated this process for all spatial bins as well for the integrated values for each galaxy in our sample. We do not discuss the ionization parameters and ISM pressures derived from \texttt{NebulaBayes} because that is outside the scope of our current work. We present the resulting metallicity maps in \cref{sec:mzgrad}.

\subsection{Metallicity gradient measurement}
\label{sec:zgrad_method}

We assigned the \oiii-flux-weighted distance of the pixels making up a given radial bin as the distance of the bin from the center. We choose \oiii for this purpose since it is typically the brightest of the observed lines. PSF-smearing has a non-negligible impact on measured metallicity gradients \citep{Yuan:2011aa, Carton:2017aa, Acharyya:2020aa, Metha:2024aa}. \citet{Metha:2024aa} outline a forward-modeling approach to account for PSF-smearing. However, the tool they provided as part of the \texttt{Lenstronomy} module \citep{Birrer:2018aa}, requires an unmasked, fully filled 2D array as the input metallicity map, whereas our metallicity maps can potentially be patchy if there are bins with no metallicity solution. Although we investigated using their tool, we concluded that the best approach for our data was to model the spatial bins directly. Thus, inspired by the prescription of \citet{Metha:2024aa}, we modeled the PSF-smeared radial gradient using Markov chain Monte Carlo \citep[MCMC;][]{Goodman:2010aa}, as follows.

We set up a simple model of an elliptical metallicity profile, with ellipticity $q$ and position angle $a$ and a smooth radial gradient characterized by slope $\nabla Z$ and central value $Z_{\rm cen}$. Then we convolved this profile with the PSF of F150W filter of \jwst/NIRISS, using the \texttt{webbpsf} python module \citep{Perrin:2014aa}. Next, we bin the smoothed metallicity profile using the spatial bins (see \cref{sec:radbin}) of the galaxy of interest, which yields the `modeled' spatially binned map. Finally, we used the \texttt{emcee} module in python \citep{Foreman-Mackey:2013aa, Foreman-Mackey:2019aa} to run MCMC sampling with 100 random walkers and 5000 iterations to get a posterior PDF of all four parameters -- $\nabla Z$, $Z_{\rm cen}$, $q$ and $a$. We report the median of the marginalized PDF for $\nabla Z$ as the measured metallicity gradient in \cref{tab:sample}, and present the 2D metallicity maps in \cref{sec:zgrad}. The discussion of the fitted morphological parameters is beyond the scope of this work, since our focus is the spatial distribution of metals. We note that although diffused ionized gas (DIG) has been shown to impact metallicity gradients \citep[e.g.,][]{Poetrodjojo:2019aa} we were unable to measure or account for it due to unreliable \sii measurements and blended \halpha in NIRISS.

\subsection{Spatially resolved SFR maps}
\label{sec:sfr}

For the 7 galaxies with measured \nii+\halpha we computed the 2-D SFR maps based on a ``corrected'' \halpha map. To correct for the \nii contribution from the observed \nii+\halpha complex, we used the 2-D metallicity map of the galaxy and the N2 (\nii/\halpha) versus metallicity calibration of \citet{Cataldi:2025aa} to determine the \nii/\halpha ratio at each spatial bin\footnote{\citet{Faisst:2018aa} provide a global correction factor for \halpha based on stellar mass and redshift. However, we consider our method of spatially resolved correction using the \citet{Cataldi:2025aa} calibration more appropriate for our study. We checked that SFRs obtained with the two correction methods differ only by $\approx 4$\%.}. Next, using this ratio, and the observed summed flux of the \nii+\halpha complex, we solved for the \halpha flux at each bin. Since the summed flux was already dereddened before this process (\cref{sec:integrated}) the ``corrected'' \halpha flux is automatically free from reddening. We then converted this ``corrected'' \halpha flux to luminosity and computed the SFR from the \halpha luminosity (L$_{H\alpha}$), as SFR (M$_{\odot}$/yr) = (7.5 $\pm$ 1.3) $\times 10^{\rm -42}$ L$_{H\alpha}$ (ergs/s) \citep{Shivaei:2015aa}.  

For the one galaxy at $z\sim3$ where \halpha is unavailable we use the following \oii-based SFR calibration from \citet{Figueira:2022aa} because their assumption of Chabrier IMF is consistent with our work: SFR (M$_{\odot}$/yr) = 10$^{-39.69 \pm 0.07} \times$ L$_{[O II]}$ (ergs/s)$^{0.96 \pm 0.01}$.

Repeating the above process for all the spatial bins we derived the 2-D SFR maps, as well as integrated SFRs. We present the results in \cref{sec:sfr_z}.

\section{Results and Interpretations}
\label{sec:results}

\subsection{Metallicity gradients}
\label{sec:zgrad}

\begin{figure*}
    \centering
    \includegraphics[width=1\linewidth, trim=1.9cm 0cm 0.cm 0.cm, clip=true]{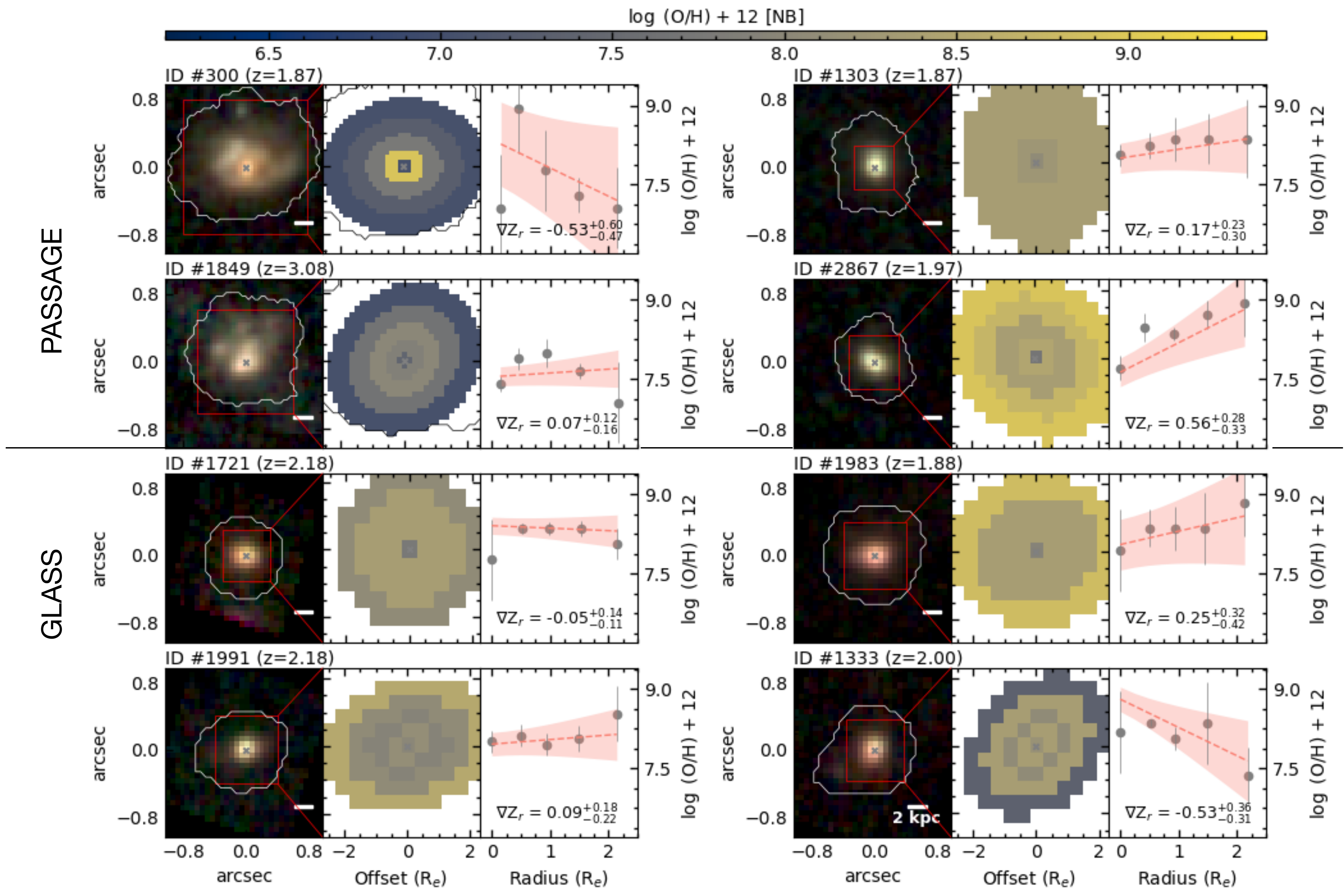}
    \caption{\texttt{NebulaBayes} metallicities for all PASSAGE (top 2 rows) and GLASS (bottom 2 rows) galaxies. In each group of plots: \textit{Left:} RGB image of the galaxy using all three NIRISS filters: \edit{F115W, F150W and F200W}. This panel is same as the top left panel of \cref{fig:galaxy}. \edit{The 2 kpc scale bar is denoted in white bar in the bottom-right.} \textit{Middle:} 2-D binned metallicity map and \textit{Right:} the corresponding radial metallicity profile. The salmon line is the \edit{best-fit} radial profile following our MCMC approach \edit{(see \cref{sec:zgrad_method} for details)}. The shaded region corresponds to that between the 16$^{\rm th}$ and 84$^{\rm th}$ percentiles of the fitted parameters. The resultant slope is annotated at the bottom. Generally, we see a scattered metallicity profile, with little-to-no azimuthal symmetry.}
    \label{fig:zgrad}
\end{figure*}

We present the \texttt{NebulaBayes} metallicity profiles and the gradients measured above in \cref{fig:zgrad}. The measured metallicity gradients for most of the galaxies are consistent with being flat within uncertainties; one galaxy \newedit{(ID \#2867 in Par028)} shows a slightly positive gradient while another \newedit{(ID \#1333 in GLASS)} hosts a negative radial gradient. Forcing a smooth linear fit to metallicity profiles with a high degree of scatter, is not necessarily optimal. Several studies have adopted non-radial approaches to quantify metallicity maps, including measuring azimuthal variations \citep[e.g.,][]{Ho:2018aa}, performing geostatistical analysis \citep{Metha:2021aa}, and characterizing the full metallicity distribution \citep{Acharyya:2025aa}. However, the quality of signal in our spatially resolved emission line maps is insufficient to employ more sophisticated approaches at this stage.

\subsection{Mass-metallicity gradient relation}
\label{sec:mzgrad}

\begin{figure*}
    \centering
    \includegraphics[width=1\linewidth]{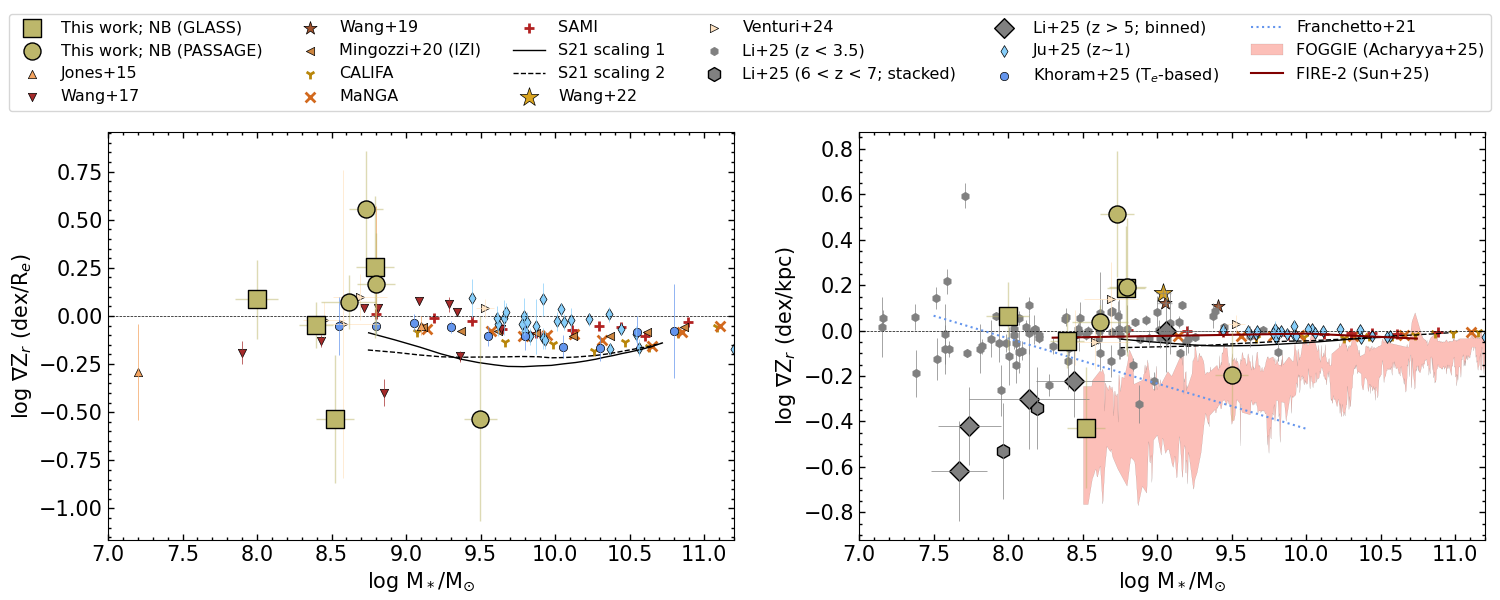}
    \caption{Mass-metallicity gradient plot for all the PASSAGE and GLASS galaxies, denoted by circles and squares respectively. The \textit{left} and \textit{right} panels correspond to gradients expressed in units of dex/R$_e$ and dex/kpc, respectively. Predictions from FOGGIE cosmological zoom-in simulations \citep{Acharyya:2025aa} are shown as the pink shaded area, and those from FIRE-2 simulations \citep{Sun:2025aa} are shown as the maroon line. The R$_e$ measurements of these predictions were unavailable, so we only plotted it on the right panel. Analytic models from \citet{Sharda:2021ab} are shown as black lines. We plot the mean gradients from low-$z$ surveys -- CALIFA \citep[triangles; ][]{Sanchez-Menguiano:2016aa}, MaNGA \citep[crosses; ][]{Belfiore:2017aa, Mingozzi:2020aa}, T$_e$-based metallicity with MaNGA \citep[blue circles; ][]{Khoram:2025aa} and SAMI \citep[plus;][]{Poetrodjojo:2018aa} -- in stellar mass bins. We adopted these data points from \citet{Sharda:2021ab} because these have been corrected for effects of spatial resolution (Acharyya et al., in prep). Measurements from several high-$z$ studies with \hst \citep{Jones:2015aa, Wang:2017aa, Wang:2019aa} and \jwst \citep{Wang:2022aa, Venturi:2024aa, Ju:2025aa, Li:2025ab} are also shown for comparison. Overall, we do not see a trend between metallicity gradient and stellar mass for our sample.}
    \label{fig:mzgrad}
\end{figure*}

We used the metallicity gradient measured from \texttt{NebulaBayes} metallicity maps to present the mass-metallicity gradient relation (MZGR) in \cref{fig:mzgrad}. Overall, we do not see any obvious trend with stellar mass. 
Our observed trend (or lack thereof) is in contrast to the FOGGIE cosmological zoom-in simulations \citep{Acharyya:2025aa}, which predict generally negative gradients at all stellar masses, with lower-mass galaxies predicted to exhibit steeper gradients. These predictions were based on tracing the evolution of 6 Milky Way-like halos from $z=4$ to $z=0.5$. In FOGGIE, this trend is primarily a time-driven evolution rather than mass-driven, and the relation with mass is largely a by- product of the fact that the galaxies build up mass over time. At higher-$z$ (lower masses) the galaxies undergo metal production but the presence of warps, misaligned disks \citep{Trapp:2026aa} and smaller disk sizes hinder metal mixing. However at lower-$z$, the galaxy disk `settles down' and metals are more readily transported away from the center, leading to flatter gradients. It appears that at high-$z$ (lower mass) the simulations are unable to reproduce the well-mixed, and therefore flatter, metallicity gradients. By comparing several modern cosmological simulations \citet{Garcia:2026aa} showed that while the redshift (or stellar mass) dependence of gradients are washed out by the large scatter at any given redshift (or stellar mass), simulations with bursty stellar feedback predict systematically flatter (by $\sim$0.2 dex) gradients than those with temporally smooth feedback prescriptions. Taken together, this could imply an insufficiently bursty stellar feedback in FOGGIE simulations at low-masses which hinders formation of flat gradients as observed in most low-mass galaxies.

Several low-$z$ studies have reported a different observational trend. \citet{Belfiore:2017aa} and \citet{Mingozzi:2020aa} have shown using MaNGA \citep{Bundy:2015aa} data a `turnover' in the mass-metallicity gradient relation around \logM $\sim10-10.5$, such that galaxies less massive than this value exhibit steeper gradients with increasing mass, and more massive galaxies exhibit shallower gradients with increasing mass. The trend at the low-mass regime was further corroborated by \citet{Li:2025ac} by studying 55 nearby dwarf galaxies. \citet{Sanchez:2017aa} used CALIFA \citep{Sanchez:2012aa} galaxies, \citet{Poetrodjojo:2018aa} used SAMI \citep{Croom:2012aa}, and \citet{Franchetto:2021aa} studied GASP \citep{Poggianti:2017aa} galaxies to report a similar turnover in the MZGR. These studies generally involved a wide range of SEL metallicity diagnostics, the choice of which has been shown to impact the measured gradient \citep{Poetrodjojo:2019aa}. However, \citet{Khoram:2025aa} confirmed this turnover by using direct (T$_e$-based) metallicity gradient measurements in a stacked MaNGA sample. \citet{Khoram:2024aa} reported negative-to-flat gradients in the star-forming disks of 11 ram-pressure stripped galaxies at $z\sim0.3$ spanning a stellar mass range of $8 \lesssim$ \logM $\lesssim 10.6$, with lower mass (\logM $<$ 9) galaxies exhibiting flat gradients. Additionally, \citet{Franchetto:2021aa} found that for a fixed stellar mass, galaxies with lower gas fraction have flatter gradients (and higher metallicity), implying they are at a more advanced stage of evolution where their gas reservoirs have been depleted and metals have been well mixed in the disk. \citet{Sharda:2021ab} used analytic models to show that the `turnover' occurs beyond a certain stellar mass when the galaxy enters an accretion-dominated phase. In this regime, increasing stellar mass leads to inflow of pristine gas to the galaxy center, thus diluting the central metallicity (i.e., flattening) the metallicity profile. However, we do not see any such `turnover' for our sample, perhaps due to our insufficient range in stellar mass, or perhaps the high-$z$ galaxies have not yet started being dominated by accretion. This is consistent with other studies at $z > 0.5$ \citep[e.g.,][]{Jones:2013ab, Jones:2015aa, Leethochawalit:2016aa, Wang:2017aa, Wang:2020aa, Wang:2022aa, Venturi:2024aa, Ju:2025aa, Li:2025ab} who have reported no trend between metallicity gradient and stellar mass in a large redshift range spanning $0.5 \lesssim z \lesssim 9$.

The blue dotted line in \cref{fig:mzgrad} corresponds to approximate ``forbidden zone'' marked by \citet{Franchetto:2021aa} to show that low-mass galaxies are unable to develop too steep gradients because they did not find any galaxy in their sample of MaNGA galaxies that lay below this demarcation. We, however, do find low mass galaxies hosting steep gradients below this limit.

Differences in metallicity calibrations aside, there exists a significant intrinsic scatter in this relation, as evidenced by the scatter and stochastic nature of the FOGGIE prediction. While stellar mass is generally a dynamical stable property tracing the cumulative effect of the galaxy's evolutionary history, metallicity gradient is an instantaneous property that is sensitive to gas flows in and around galaxies. As such, gradients can undergo significant changes on short time scales ($\sim$50 Myr), as seen in FOGGIE galaxies in \cref{fig:mzgrad}. This increases the intrinsic scatter in the gradients measured for a given population of galaxies. Indeed, observational studies of large samples \citep[e.g.,][]{Belfiore:2017aa, Sanchez-Menguiano:2018aa} show $\sim0.3-0.5$ dex/kpc scatter. The CALIFA, MaNGA and SAMI data plotted in \cref{fig:mzgrad} represent the mean behavior, and therefore exhibit a gentle variation with mass. Given our small sample size, future \jwst studies with a larger sample across a wider mass range ($\lesssim$ 10$^{11}$\Msun) are required, in order to confirm any mass–gradient relation at high-$z$. 

\subsection{SFR surface density-metallicity correlations}
\label{sec:sfr_z}

\begin{figure*}
    \centering
    \includegraphics[width=1\linewidth, trim=20cm 0cm 0.cm 0.cm, clip=true]{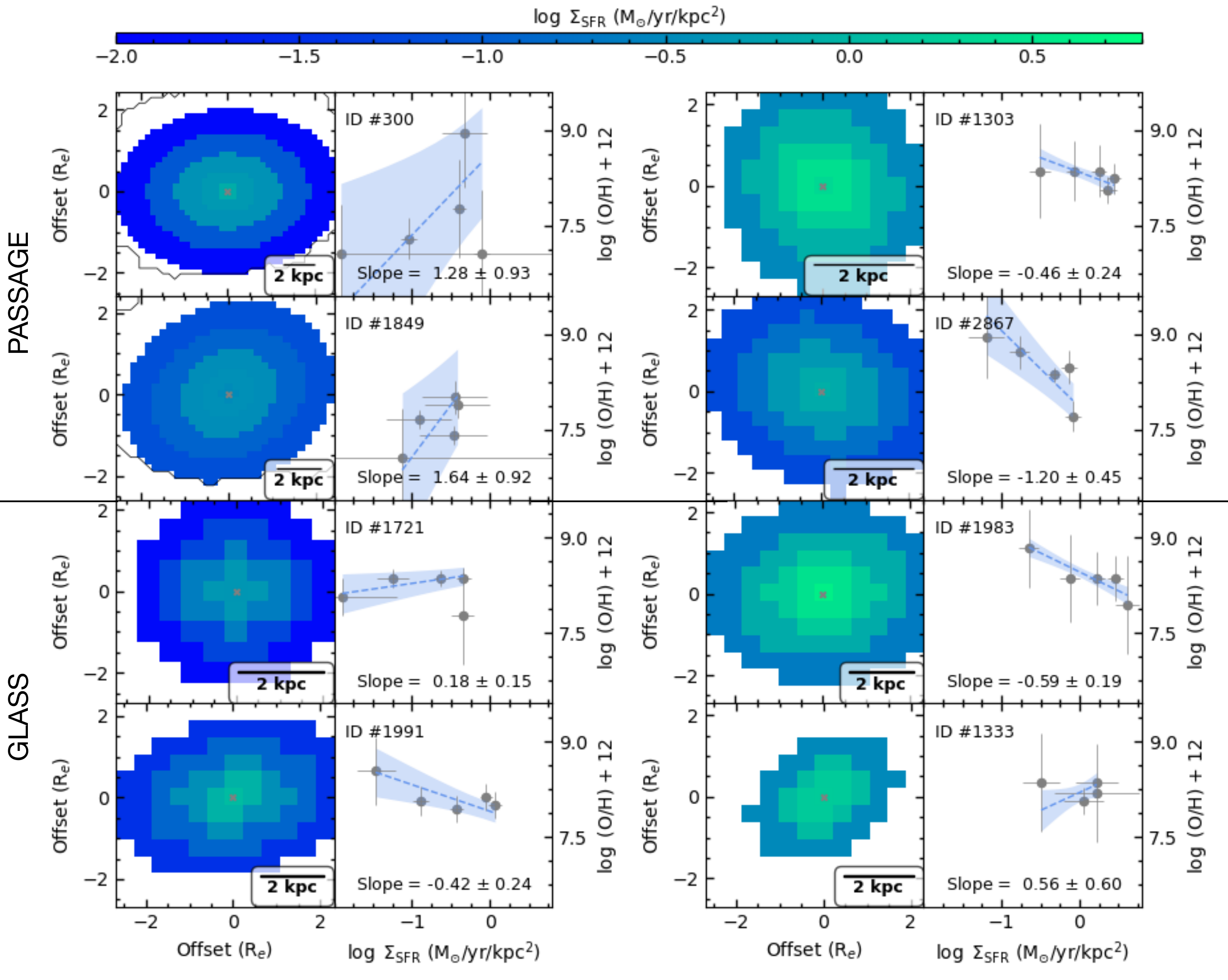}    
    \caption{2-D SFR maps and their correlations with metallicities for all PASSAGE (top 2 rows) and GLASS (bottom 2 rows) galaxies. In each pair of panels: \textit{Left:} SFR surface density ($\Sigma_{\rm SFR}$) map of the galaxy derived from the corrected \halpha map, with the exception of galaxy ID \#1849 where SFR was measured from the \oii map (\cref{sec:sfr}), and \textit{Right:} $\Sigma_{\rm SFR}$ vs the \texttt{NebulaBayes} metallicity estimated for each spatial bin. While the relation has a significant scatter, fitting with the \edit{Orthogonal Distance Regression} (ODR) method generally yields a diverse range of correlations (\cref{sec:sfr_z}). \edit{The fit is denoted by the dashed blue line and the slope is annotated at the bottom of each panel. The shaded region corresponds to the 1-$\sigma$ uncertainty of the fitted profile, derived by propagating the uncertainties of the linear fit parameters.} This figure highlights the diversity of the local metallicity-$\Sigma_{\rm SFR}$ slopes seen in our sample.}
    \label{fig:zsfr}
\end{figure*}

The Fundamental Metallicity Relation (FMR) -- a correlation between global measurements of galaxy metallicity, stellar mass and SFR -- has been widely studied in order to understand the interplay of these factors at the global scale \citep[e.g., ][]{Mannucci:2010fk, Maiolino:2019aa, Curti:2020ab, Nakajima:2022aa}. Recent years have witnessed a growing interest in investigating these relations at the local ($\sim1$ kpc) scales -- to determine whether the local physics drives the global evolution of the galaxy or vice-versa. Correlating local metallicity, SFR surface density and stellar (or gas) mass surface density is crucial to address the above question. Although we do not have the third quantity in a spatially resolved sense\footnote{The available spatially resolved photometry for our sample has insufficient S/N for reliable spatially resolved SED fitting.}, we investigate correlations between the first two quantities.

\cref{fig:zsfr} shows the correlation between the spatially resolved SFR surface density ($\Sigma_{\rm SFR}$) maps with the metallicity maps for our sample. We adopted the Orthogonal Distance Regression (ODR) method of the \texttt{scipy} module in Python to fit a linear model to the metallicity-vs-$\Sigma_{\rm SFR}$ plot. We see a diverse range of metallicity-$\Sigma_{\rm SFR}$ slopes, ranging from strongly positive to weakly negative. All our galaxies show centrally dominated star-formation, consistent with inside-out growth scenario. The metals however are comparatively more spread-out to larger radii (\cref{fig:zgrad}), perhaps due to strong radial mixing, which drives the diversity in metallicity-$\Sigma_{\rm SFR}$ slopes. 

\citet{Sanchez-Menguiano:2019aa} and \citet{Teklu:2020aa} report metallicity-$\Sigma_{\rm SFR}$ slopes for individual MaNGA ($z\lesssim0.15$) galaxies ranging from weakly positive to weakly negative, the majority being consistent with a flat slope, whereas \citet{Sanchez-Almeida:2018aa} reported negative-to-flat slopes for a sample of 14 dwarf galaxies. This diversity in individual galaxy behavior agrees with our results. However, upon combining a large sample ($\sim$600-2000) of star-forming MaNGA galaxies, studies have shown a weak but non-zero anti-correlation between local metallicity and SFR surface density, for a given stellar mass \citep{Barrera-Ballesteros:2016aa, Maiolino:2019aa, Baker:2023aa}. The anti-correlation disappears at larger stellar masses and has been interpreted as a signature of infalling metal poor gas, which spurs new SF and dilutes local metallicity \citep{Maiolino:2019aa}, or due to self-enrichment via stellar winds and supernovae \citep{Sanchez-Almeida:2018aa}. Other studies have indirectly discussed proxies of the local metallicity-$\Sigma_{\rm SFR}$ relation via specific-SFR \citep{Yao:2022aa} or gas fraction \citep{Cresci:2010aa, Trayford:2019aa, Franchetto:2021aa}. They all agree on a weak anti-correlation between metallicity and SFR surface density. While our findings of diverse metallicity-$\Sigma_{\rm SFR}$ slopes at $z\sim2$ are in contrast with the low-$z$ ($z\lesssim 0.1$) studies, but they agree in terms of the galaxy-to-galaxy scatter. Although our small sample does not provide a statistically significant comparison to large local surveys with (e.g., MaNGA), this might be hinting at a true redshift-evolution of physical processes.

\subsection{Implications of the SFR-metallicity correlation}
\label{sec:mzsfr}

\begin{figure}
    \centering
    \includegraphics[width=1\linewidth]{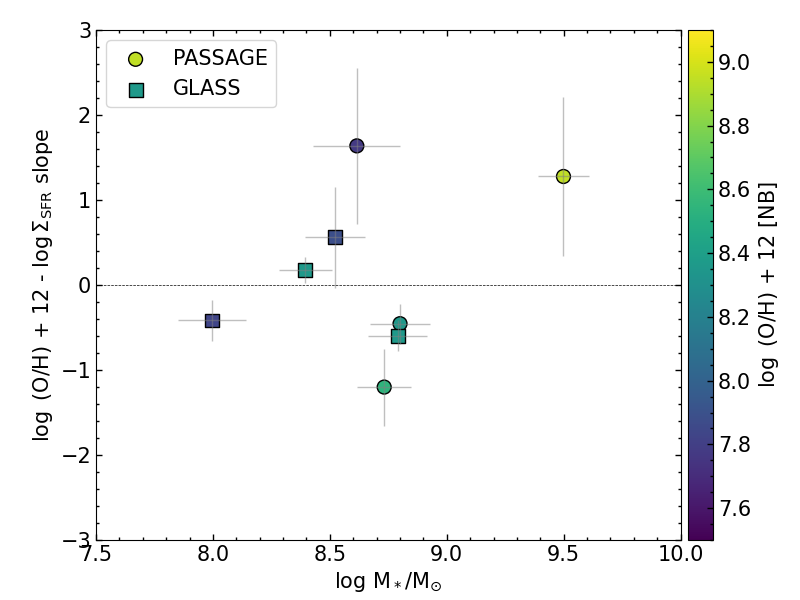}
    \caption{Slope of the metallicity vs $\Sigma_{\rm SFR}$ relation as a function of stellar mass  for all the PASSAGE (circles) and GLASS (squares) galaxies. The points are color-coded by the integrated metallicity obtained from \texttt{NebulaBayes}. A weak positive correlation between metallicity-$\Sigma_{\rm SFR}$ slope and stellar mass is seen in our sample.}
    \label{fig:mzsfr}
\end{figure}

To quantify the impact of global stellar mass on the local correlation between metallicity and SFR surface density, we plot the slopes measured in \cref{fig:zsfr} against the stellar masses of each galaxy, in \cref{fig:mzsfr}. A hint of a weak correlation emerges, such that local metallicity and $\Sigma_{\rm SFR}$ are more strongly correlated for more massive galaxies, but it is difficult to draw strong conclusions due to a lack of sufficiently large range of stellar masses in our sample. This is in agreement with \citet{Sanchez-Menguiano:2019aa}, who reported a positive correlation between stellar mass and metallicity--$\Sigma_{\rm SFR}$ slopes (after subtracting the radial profiles). For the galaxies where the metallicity and $\Sigma_{\rm SFR}$ are positively correlated, the two quantities become increasingly independent as the stellar mass decreases, i.e., the metallicity-$\Sigma_{\rm SFR}$ slope approaches zero. However, we also measure negative metallicity-$\Sigma_{\rm SFR}$ slopes (i.e. anti-correlation) for 50\% of our galaxies, implying a significant diversity in the physical processes at high-$z$. While the precise values of the slopes are sensitive to the spatial binning scheme and metallicity diagnostics used, the diversity as well as the weak correlation with stellar mass is robust.

\section{Discussion: Metal-mixing timescale}
\label{sec:disc}

Broadly, two competing time-scales shape the metallicity distribution: (a) metal production time-scale, which is directly related to $\Sigma_{\rm SFR}$, and (b) metal mixing time-scale, which is dictated by the strength of stellar feedback, along with other physical processes \citep{Sharda:2021aa, Sharda:2024aa}. Although massive galaxies might produce more powerful feedback \citep[e.g.,][]{Tremonti:2004aa, Recchi:2013aa} modes, they have a stronger gravitational potential too, hindering gas- (and therefore, metal-) transport. Low-mass galaxies on the other hand have been known to promote rapid metal re-distribution owing to their weaker potential well \citep[e.g.,][]{Rupke:2010ab, Chisholm:2018aa}. Therefore, the metallicity-$\Sigma_{\rm SFR}$ slope can hold clues to metal-mixing timescales ($t_{mix}$).

In this work we present a novel framework to relate the metallicity($Z$)-$\Sigma_{\rm SFR}$ slope to $t_{mix}$ as

\begin{equation}
\label{eq:tmix}
    \boxed{\frac{d \log Z}{d\log \Sigma_{\rm SFR}} = \beta \frac{\epsilon}{1 - \epsilon}\quad \text{where} \quad \epsilon = t_{mix}B\Sigma_{\rm SFR}^\beta}
\end{equation}

\noindent where $B=0.004$ and $\beta=0.333$ are constants obtained from the Kennicutt-Schmidt law \citep{Kennicutt:1998ab, Kennicutt:2021aa} $\Sigma_{\rm SFR} = A \Sigma_g^\alpha$, such that $B=A^{1/\alpha}$ and $\beta = 1 - 1/\alpha$. A full derivation of this relation is given in \cref{sec:ap_tmix}, and we focus here on its implications for our results.

\cref{eq:tmix} relates the $\frac{d \log Z}{d\log \Sigma_{\rm SFR}}$ slope directly to metal mixing timescale $t_{mix}$ and $\Sigma_{\rm SFR}$. Moreover, it shows that the metallicity-$\Sigma_{\rm SFR}$ slope is monotonically, albeit non-linearly, related to both $t_{mix}$ and $\Sigma_{\rm SFR}$. For a constant $\Sigma_{\rm SFR}$, a larger $t_{mix}$ (metals are less mixed) yields a larger metallicity-$\Sigma_{\rm SFR}$ slope (efficient build up of the metals near their production site), and vice versa. Similarly, for a constant $t_{mix}$, a higher $\Sigma_{\rm SFR}$ yields a larger metallicity-$\Sigma_{\rm SFR}$ slope. Therefore, both a high $t_{mix}$ and a high stellar-mass can independently cause a large metallicity-$\Sigma_{\rm SFR}$ slope. In order to isolate the effect of $t_{mix}$, we need to account for $\Sigma_{\rm SFR}$ in \cref{eq:tmix}, which we did by substituting $\Sigma_{\rm SFR}$ with the mean $\Sigma_{\rm SFR}$ for each galaxy.

\begin{figure}
    \centering
    \includegraphics[width=1\linewidth]{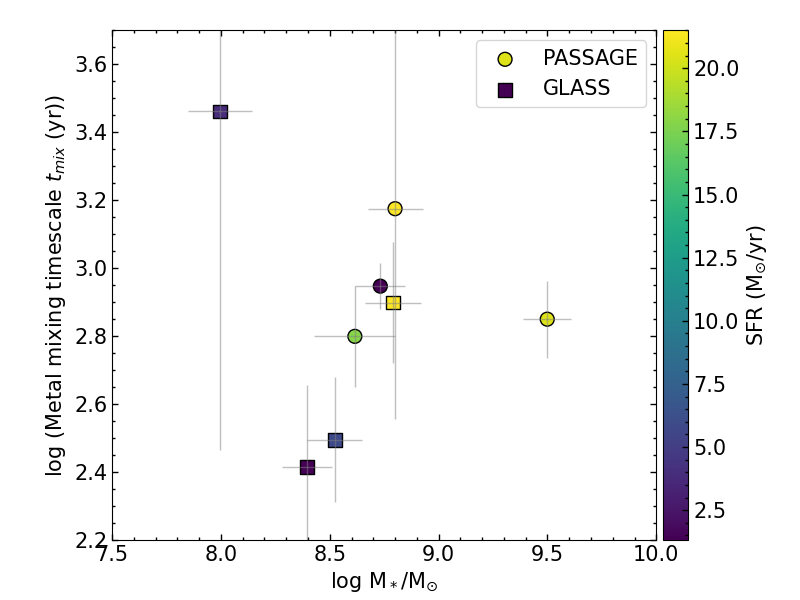}
    \caption{Metal mixing timescale $t_{mix}$ (\cref{sec:disc}) as a function of stellar mass for PASSAGE (circles) and GLASS (squares) galaxies. The points are color-coded by the integrated SFR. A weak positive correlation is seen between the mixing timescale and stellar mass for our sample.}
    \label{fig:mtmix}
\end{figure}

We show $t_{mix}$ as a function of stellar mass in \cref{fig:mtmix}, emphasizing that $t_{mix}$ may not represent true physical mixing time. The $t_{mix}$ might better be viewed as an effective timescale combining multiple processes (diffusion, advection, feedback-driven outflows). Indeed the typical dynamical timescales for these galaxies, computed using the stellar masses and R$_e$, are $\sim$10\,Myr which is significantly longer than the typical $t_{mix}\sim10^3$\,yr. As such, our model might be capturing an effective timescale averaged over small scales where mixing is fast, rather than the global mixing timescale. We see a very weak ($\sim 0.1\sigma$) trend of higher mixing timescales with increasing stellar mass, albeit this trend is consistent with being flat within uncertainties. The significance value is based on the slope $= 22.56 \pm 280.61$ we computed for \cref{fig:mtmix} (but not shown in the Figure). Note that omitting the data-point with significantly large uncertainties in $t_{mix}$ (the data point at \logM $\sim$8) from this calculation, yields a slope of $0.96 \pm 0.6$. Although it is difficult to draw strong conclusions based on our limited sample size and the caveats of the measurements discussed in \cref{sec:zdiag_sel}, this simple theoretical framework linking metallicity-$\Sigma_{\rm SFR}$ slope to $t_{mix}$ will be useful for future studies with larger and more robust samples.

\section{Summary and Conclusions}
\label{sec:sum}
In this work we demonstrated the scientific utility of NIRISS/WFSS data at an individual galaxy scale. We applied state-of-the-art models and techniques to the best available NIRISS galaxy data from the PASSAGE and GLASS programs, to understand the role of stellar feedback on the chemical evolution of galaxies. Although it is difficult to draw strong science conclusions based on the limited sample size, the overall exercise: (a) demonstrates the challenges and potential of such studies with NIRISS/WFSS data;  and (b) sets up the framework aimed at discovering new science once a larger and more robust sample is available. Specifically, we presented spatially resolved metallicity gradient measurements with NIRISS/WFSS observations for eight galaxies, increasing the sample size of galaxies with such measurements by an order of magnitude\footnote{\citet{Wang:2022aa} was the first study to present metallicity map for one galaxy studied with NIRISS/WFSS.}. We also presented the SFR maps, and investigated the correlations between spatially resolved SFR and metallicity. We summarize the lessons learned from this work as follows.

\begin{itemize}
    \item The \sii line map observed with NIRISS/WFSS was unreliable. However, the \neiii and \oii maps, if detected, can help identify the dominant photoionization mechanism in a spatially resolved way, although we were able to use it only for  integrated measurements due to insufficient signal in the \neiii maps. We presented a new SF-demarcation line in the \oiii/\hb vs \neiii/\oii  parameter space, based on publicly available MAPPINGS v5.1 \hii region \newedit{AGN} model grids \citep{Thomas:2018aa}. \newedit{We caution the reader that these two sets of grids overlap partially in the \ohno parameter space, such that galaxies that lie below our proposed demarcation maybe AGN or SF-dominated, but those that lie above are inconsistent with photoionization by SF.}
    \item Having measured mostly flat or slightly positive spatial metallicity gradients, we showed on the mass-metallicity gradient plot (\cref{fig:mzgrad}), a broad agreement with existing observations in terms of the observed scatter and the range of gradients occupied, no visible trend with stellar mass, nor a turnover, as reported in multiple studies at low-$z$ ($z\lesssim1$). The lack of trend between stellar mass and metallicity gradient can be hinting at the fact that the high-$z$ galaxies have not yet started being dominated by accretion. Like most observed gradients, our results are in contrast with the trends predicted by simulations.
    \item Comparing local metallicity and star-formation rate surface density, for the first time at $z\gtrsim2$, we saw a weak trend in the metallicity-$\Sigma_{\rm SFR}$ slope and global stellar mass in \cref{fig:mzsfr}. The trend with stellar mass became clearer upon interpreting this as an indirect probe of the effective timescale for metal-mixing ($t_{mix}$) induced by stellar feedback in galaxies. Although we find a weak trend of higher $t_{mix}$ with increasing stellar mass, it is challenging to draw strong conclusions given our small sample size, and limited stellar mass range coverage. 
\end{itemize}

Although we analyzed a small sample size, our work demonstrates the potential of NIRISS/WFSS observations in addressing key questions in galaxy evolution. This paves the way for future studies with larger samples of individual galaxies.

\section*{Data availability}
The data used are available at the MAST, with JWST programme ID 1527 (for PASSAGE) and ID 1324 (for GLASS).

\begin{acknowledgements}
We thank the anonymous referees for their thorough and positive feedback, which has improved the manuscript. We thank Giacomo Venturi for productive discussions related to metallicity diagnostics, and Zihao Li for sharing their data via private communication. This research was supported by the International Space Science Institute (ISSI) in Bern, through ISSI International Team project "Bringing PASSAGEers together from around the world to solve the Epoch of Reionization" (ISSI Team project \#24-624). AA, BV, PW, GR and AG acknowledge support from the INAF Large Grant 2022 “Extragalactic Surveys with JWST” (PI Pentericci) and from the European Union – NextGenerationEU RFF M4C2 1.1 PRIN 2022 project 2022ZSL4BL INSIGHT. PW and BV acknowledge support from the INAF Mini Grant ``1.05.24.07.01 RSN1: Spatially-Resolved Near-IR Emission of Intermediate-Redshift Jellyfish Galaxies'' (PI Watson). AJB acknowledges funding from the “FirstGalaxies” Advanced Grant from the European Research Council (ERC) under the European Union’s Horizon 2020 research and innovation program (Grant agreement No. 789056). HA acknowledges support from CNES, focused on the JWST mission, and the Programme National Cosmology and Galaxies (PNCG) of CNRS/INSU with INP and IN2P3, co-funded by CEA and CNES. XW is supported by the National Natural Science Foundation of China (grant 12373009), the CAS Project for Young Scientists in Basic Research Grant No. YSBR-063, the Fundamental Research Funds for the Central Universities, the Xiaomi Young Talents Program, and the China Manned Space Program with grant no. CMS-CSST-2025-A06. The data were obtained from the Mikulski Archive for Space Telescopes (MAST) at the Space Telescope Science Institute, which is operated by the Association of Universities for Research in Astronomy, Inc., under NASA contract NAS 5-03127 for JWST. The observations used in this work are associated with program IDs 1571 (PASSAGE) and JWST-ERS-1324 (GLASS) and we acknowledge financial support through these respective grants. The python packages {\sc matplotlib} \citep{matplotlib2007}, {\sc numpy} \citep{numpy2011}, \textsc{scipy} \citep{scipy2020}, {\sc yt} \citep{yt2011}, {\sc datashader} \citep{datashader2022}, and {\sc Astropy} \citep{astropy2013,astropy2018,astropy2022} were all used in parts of this analysis.

\end{acknowledgements}

\bibliographystyle{aa} 
\bibliography{paper}

\begin{appendix}
\renewcommand\thefigure{\thesection\arabic{figure}}  
\setcounter{figure}{0}

\section{Galaxy size vs spatial resolution}
\label{sec:ap_re}

\begin{figure}
    \centering
    \includegraphics[width=1\linewidth]{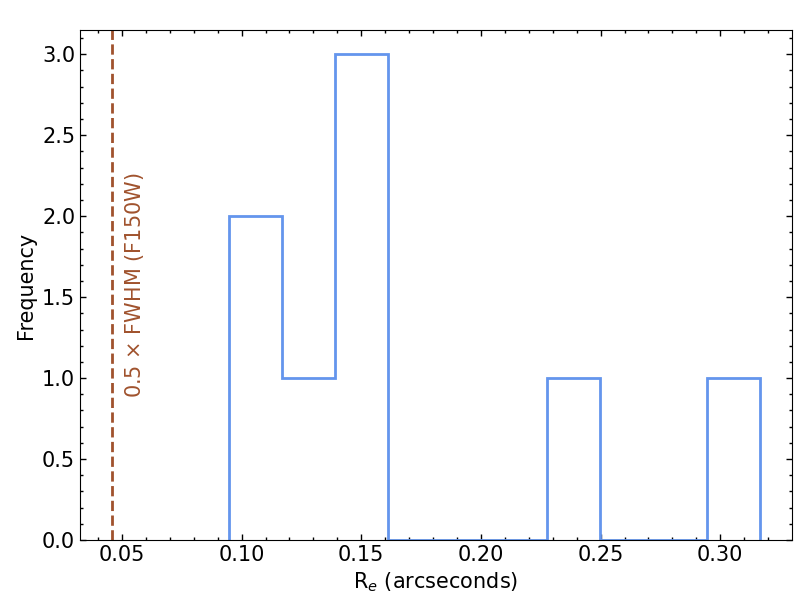}
    \caption{Distribution of the effective radii measured for our sample (see \autoref{sec:re}) is shown in blue. The brown vertical line denotes half the FWHM of the PSF in the F150W NIRISS filter. This demonstrates that all our galaxies are larger than the FWHM.}
    \label{fig:re}
\end{figure}

\autoref{fig:re} shows the distribution of R$_e$ for our sample, along with the half the FWHM of the PSF in the F150W filter. Each of our galaxy spans at least two FWHM of PSF, in diameter, demonstrating that they are resolved sources.

\section{Choice of integrated line fluxes}
\label{sec:ap_int}

\begin{figure*}
    \centering
    \includegraphics[width=1\linewidth]{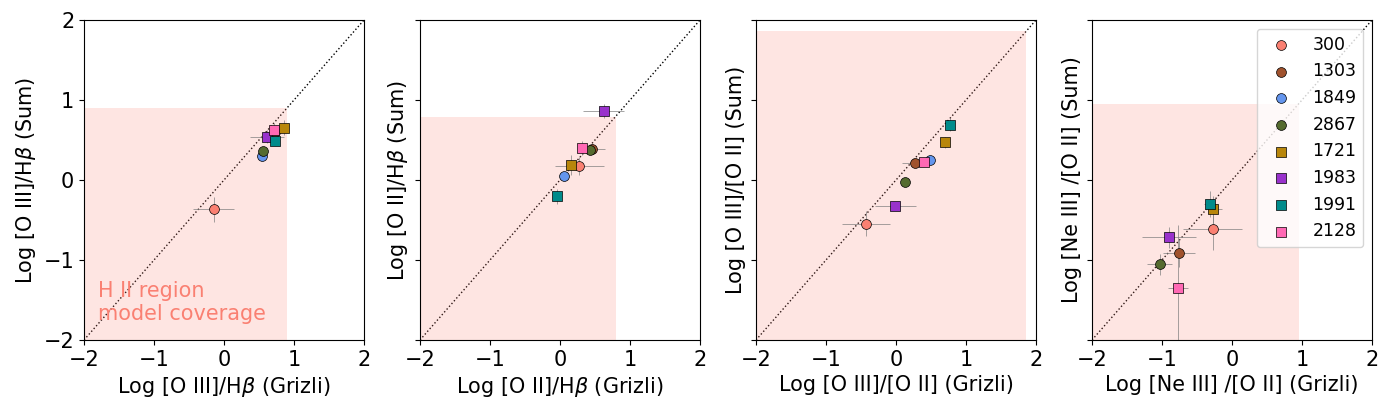}
    \includegraphics[width=1\linewidth]    {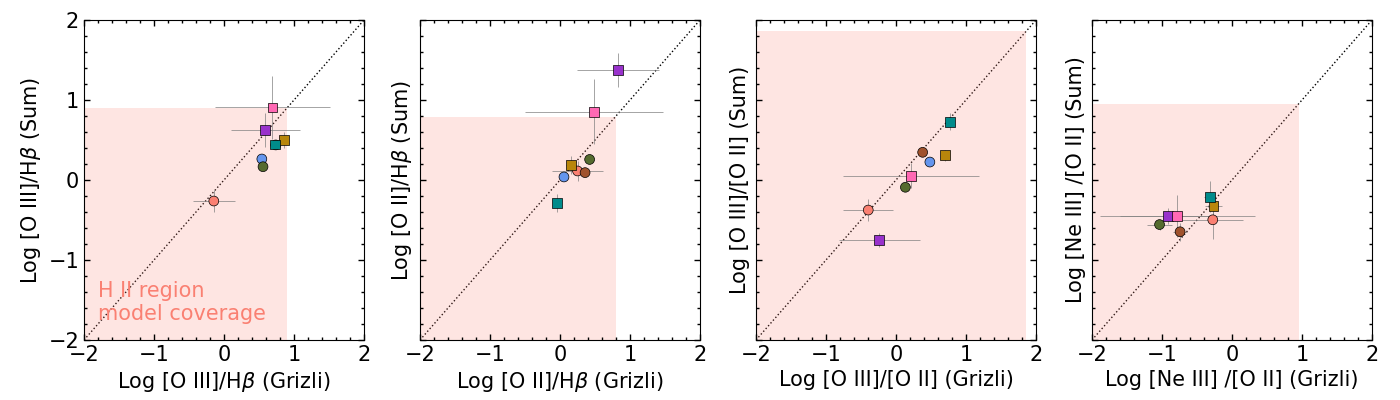}
    \caption{Comparison of flux ratios obtained from \grizli-reported integrated fluxes and from summing up the 2D flux maps. \textit{Top} and \textit{bottom} panels correspond to the summing up of fluxes performed within $\pm$2.5 R$_e$ and within the full segmentation map, respectively. In each row, from left to right, we plot the \oiii/\hb \oii/\hb, \oiii/\oii and \neiii/\oii ratios, because these are the ratios most relevant for metallicity estimation and for determining the dominant ionization mechanism. The colors correspond to different galaxies in our sample. There is a general agreement between the two within $\sim25\%$ for most galaxies, and no systematic offset. The salmon shaded region in each panel denotes the extent of the ratios reproduced by the MAPPINGS v5.1 \hii region models \citep{Thomas:2018aa}. It is challenging to estimate metallicities of galaxies with ratios outside this region (see \cref{sec:outside_model}).}
    \label{fig:line_ratio_comp}
\end{figure*}

\cref{fig:line_ratio_comp} shows the \grizli-reported integrated line flux ratios against those obtained by summing up the 2D line maps, with circles and squares denoting PASSAGE and GLASS-NIRISS galaxies respectively. We did not see any systematic offset between the two methods of estimating integrated line flux ratios, and they generally agree with each other within $\sim$0.2 dex for our galaxies. As discussed in \cref{sec:integrated}, we chose to use the summed flux of the 2D line maps as the integrated line flux for each galaxy for all our analysis.

In each panel of \cref{fig:line_ratio_comp}, the salmon colored background denotes the parameter space covered by the MAPPINGS v5.1 photoionization model grids. These model grids are used for metallicity estimation employing a Bayesian approach, discussed in \cref{sec:zdiag_nb}. Integrated \oiii/\hb and \oii/\hb ratios of a few galaxies in our sample lie slightly outside the model coverage. This can potentially impact the metallicity estimation using these ratios. We discuss this in more detail in \cref{sec:outside_model} and \cref{sec:zdiag_nb}.

\section{Mass-excitation diagram}
\label{sec:ap_mex}

\begin{figure}
    \centering
    \includegraphics[width=1\linewidth]{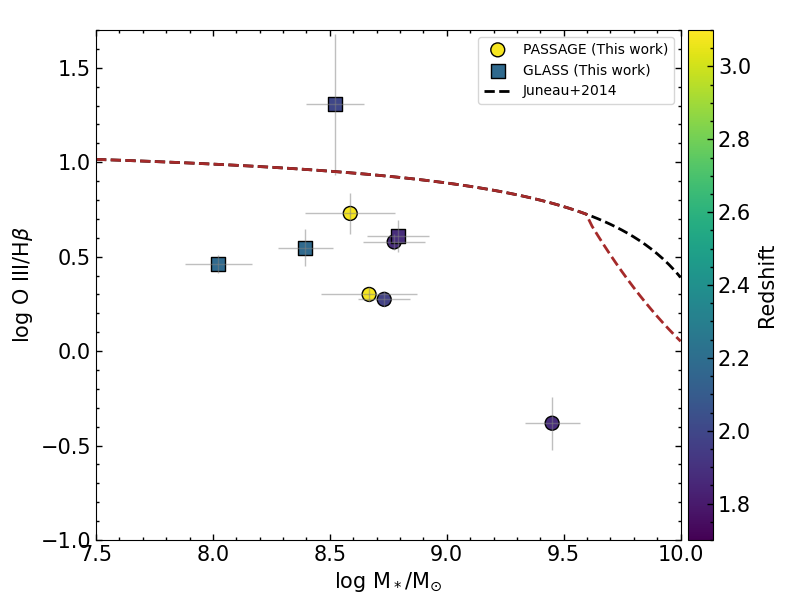}
    \caption{Mass-excitation (MEx) diagram for the global properties of the PASSAGE and GLASS-NIRISS galaxies depicted with circles and squares respectively. The lower envelope of the \citet{Juneau:2014aa} line demarcates purely star-forming galaxies.}
    \label{fig:mex}
\end{figure}

The mass-excitation (MEx) diagram---a parameter space of the \oiii/\hb ratio and the stellar mass---has been increasingly commonly used to dissociate AGN-hosts from SF galaxies at high-$z$ \cite[e.g.,][]{Juneau:2013aa, Henry:2013aa, Henry:2021aa}. \cref{fig:mex} shows the MEx diagram for our sample, which is at $z\sim2-3$. Details of the stellar mass measurements are in \cref{sec:mstar}. 
Galaxies that lie below the red dashed line (i.e., the lower envelope of the \citealt{Juneau:2014aa} lines) have $<20\%$ probability of being AGN-hosts, while those above the upper envelope (black dashed line) have $>80\%$ probability of hosting an AGN. \cref{fig:mex} confirms that all but one of our galaxies are consistent with being star-forming. 

This one GLASS galaxy, although consistent with being star-forming within error bars, has large uncertainties in the line ratio. This is due to an over-subtraction issue in the \hb map, that leads to unusually low integrated \halpha\footnote{Although we do not show it in the paper, this skews the spatially resolved \hbeta, and consequently the metallicity maps, as we noticed in our internal investigations. Hence we omitted this object.} accompanied with large uncertainties. For the sake of obtaining a ``safer'' sample, we discarded this galaxy from our sample.

\section{Blending of \neiii with \heilam}
\label{sec:ap_hei}

\begin{figure}
    \centering
    \includegraphics[width=0.95\linewidth]{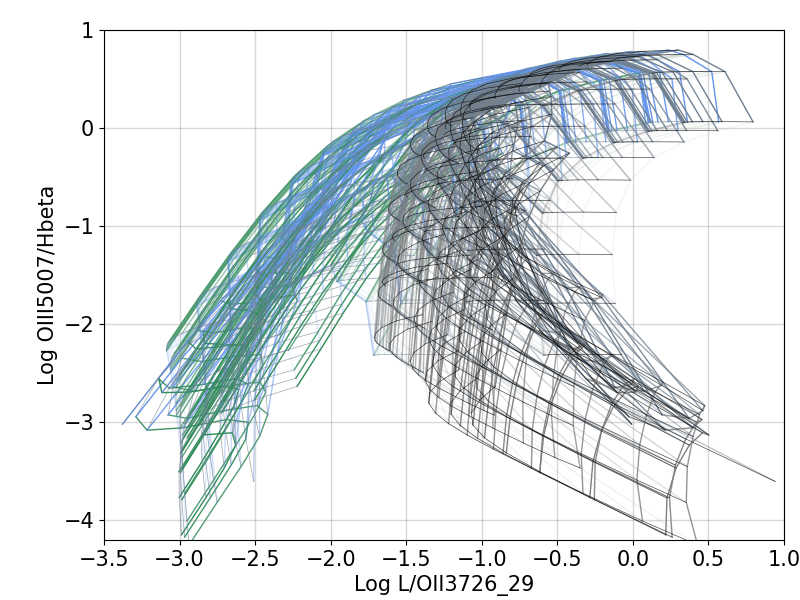}
    \caption{Same as left panel of \cref{fig:nb_fit}, but now the numerator used to compute the ratio on the x-axis corresponds to L=\neiiilam for the colored grid (same as the original figure) and L=\neiiilam + \heilam for the gray grid.}
    \label{fig:nb_hei}
\end{figure}

Although \neiiilam is blended with \heilam at NIRISS spectral resolution, we did not perform any correction to obtain the intrinsic \neiii for our analysis, because the correction factor depends on the physical properties of the galaxy in a degenerate fashion. To demonstrate the impact of this blending on the SF-AGN demarcation we presented in \cref{sec:ohno}, in \cref{fig:nb_hei} we compare the NebulaBayes photoniozation model grids with and without adding the \heilam component to \neiiilam (gray and colored grids, respectively). While the grids unsurprisingly differ from each other on the whole, their upper envelopes at high values of ratios (majority of our sample has \oiii\hb $\gtrsim$ 0.3 and \neiii/\oii $\gtrsim$ -1) are similar. Therefore the demarcation we laid out for purely SF galaxies still hold true even if the \neiii fluxes have some \heilam component in them.

\section{Strong line metallicity diagnostics}
\label{sec:zdiag_sel}

We employed the R2 (\oiilam/\hb), R3 (\oiiilam/\hb), and R23 ((\oiilam + \oiiiablam)/\hb) diagnostics from \citet[][hereafter \cataldi]{Cataldi:2025aa} on each spatial bin. Although the \citet{Curti:2020ab} calibrations are commonly used by the community, including for intermediate-$z$ ($1 \lesssim z \lesssim 2$) studies \citep[e.g.,][]{Revalski:2024aa}, these calibrations are mostly derived from local galaxies. While \citet{Revalski:2024aa} recently validated these calibrations out to $z\sim2.5$, the \cataldi calibrations are based on direct (T$_e$)-method metallicity measurements of $\sim$100 high-$z$ ($2 \lesssim z \lesssim 3$) galaxies, and are therefore better suited for our sample. Blending of \nii and \ha in the NIRISS spectra prevents reliable use of diagnostics involving either line separately, ruling out the recent AGN and SF+AGN metallicity calibrations by \citet{Peluso:2025aa}, which require them to be resolved.
 
\subsection{Choice of the metallicity branch}
\label{sec:double_valued}

\begin{figure*}
    \centering

    \begin{subfigure}{\linewidth}
        \centering
        \includegraphics[width=0.8\linewidth]{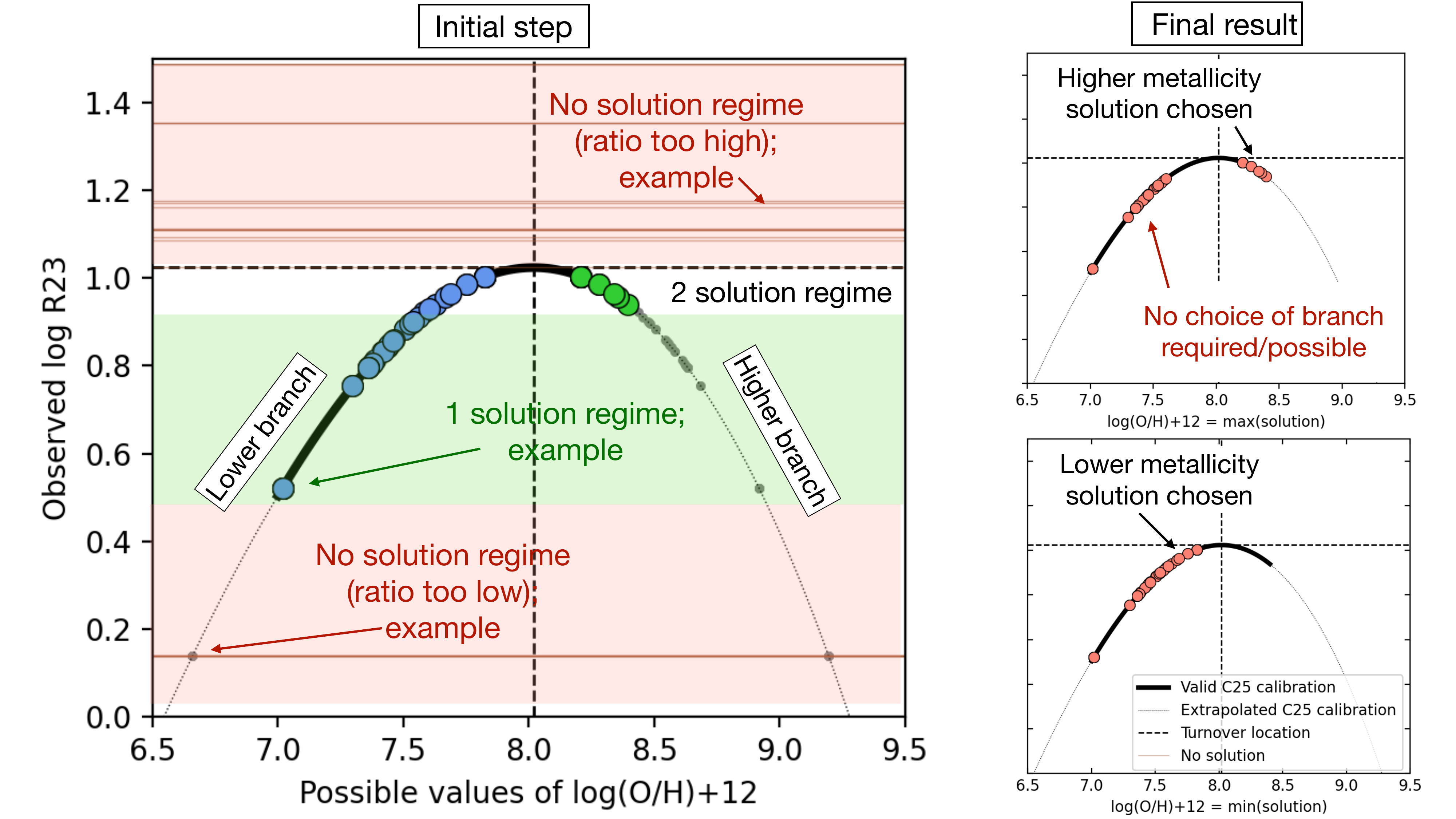}
        \caption{R23 calibration from \cataldi applied to our example galaxy. The left panel indicates the first step---observed line ratios of each bin vs the potential solutions for metallicity as colored circles. The solid curve is the \cataldi R23 calibration within the metallicity range for which it is valid, and the dotted curve is the extrapolation which is unreliable. Observed ratios that do not overlap with the calibration's range are shown as brown horizontal lines. The gray circles, denoting where the observed ratio coincides with the extrapolated calibration, are discarded from our measurements. The shaded areas denote scenarios where one (green), two (white) or no (pink) solution is available for a given observed line ratio. The right panel shows the final result, corresponding to the higher (top) and lower (bottom) metallicity solutions chosen wherever available.}
        \label{fig:zdiag_branch_example}   
    \end{subfigure}
    \vspace{1em}
    \begin{subfigure}{\linewidth}
        \centering
        \includegraphics[width=0.9\linewidth]{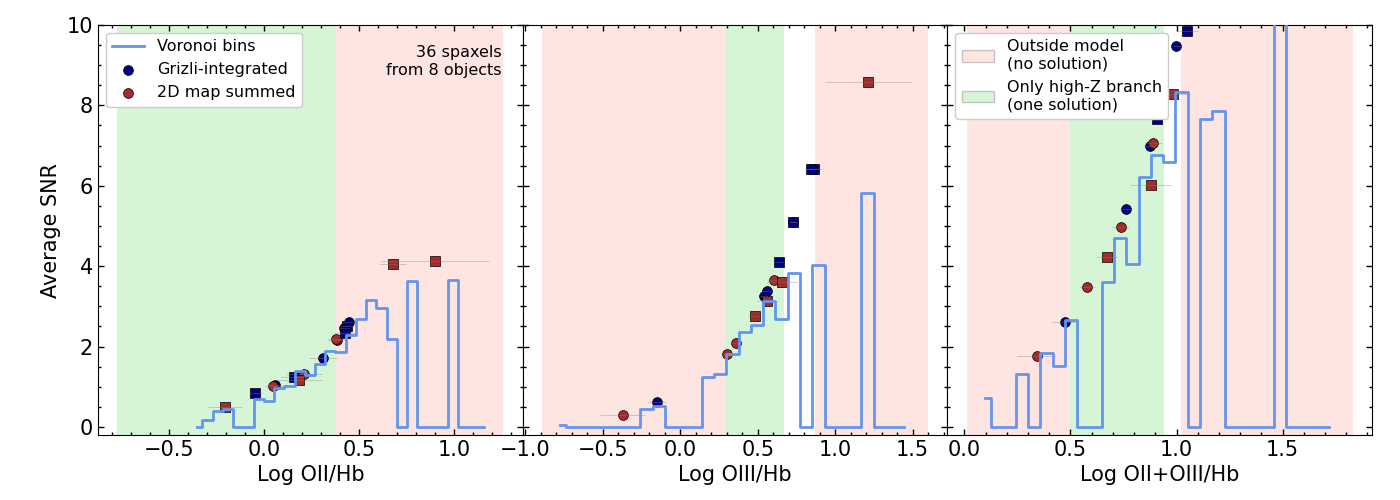}
        \caption{Distribution of the spatial binned line ratios for all spatial bins of the galaxies in our sample, as a function of the mean S/N (of the ratio), for the R2 (\textit{left}), R3 (\textit{middle}) and R23 (\textit{right}) ratios. The global measurements of line ratios---both the \grizli-reported value (navy blue) and the summed value that we have used in this analysis (brown)---are shown as circles and squares for PASSAGE and GLASS galaxies, respectively. The vertical shaded areas represent various regimes where single, double, or no solutions are available for metallicity, based on the \cataldi calibrations (see \cref{sec:double_valued}). Note that the \oii/\hbeta calibration does not have a two-solution regime because it has a linear relation with metallicity.}
        \label{fig:line_ratio_hist}
    \end{subfigure}
    \caption{Demonstration of the impact of the double-valued nature of strong emission line diagnostics.}
\end{figure*}

As shown in \cref{fig:zdiag_branch_example}, the \cataldi diagnostics used here are double-valued in metallicity, with a turnover ranging between $8 < \logOH < 8.5$. This is because the Oxygen emission lines are produced by collisional excitation. A very low abundance of \oiii or \oii relative to \hb is only possible due to insufficient avenues for collisional excitation, implying a low temperature in the nebula, which in turn implies the presence of a large amount of metals that helps cool the gas via radiative cooling channels. Conversely, a high \oiii/\hb or \oii/\hb ratio could be either due overall high abundance of Oxygen (i.e., high metallicity) or a low abundance of Oxygen, leading to insufficient cooling and consequently higher chance of collisional excitation events.

Typically such degeneracies are resolved using an independent line ratio that varies monotonically with metallicity, such as \nii/\ha \citep[e.g.,][]{Kewley:2002fk}. Unfortunately, the blending of \halpha with \nii, due to R$\sim$150 resolution of NIRISS, prevents us from adopting this strategy. Usually studies simply adopt one of the metallicity branches \citep[e.g.,][]{Maiolino:2008aa, Henry:2021aa}. This choice is often made based on one of the following arguments: (a) global stellar mass serves as a proxy for metallicity and therefore can provide some indication of the expected metallicity, (b) non-detection of $\left[\ion{O}{iii}\right]\,{\lambda 4363}$ emission implying a low electron temperature ($T_e$) and therefore higher metallicity, (c) photoionization models are more sensitive in the higher branch, while empirical calibrations ($T_e$ based) are preferred for the lower branch.

For illustration, we show observed binned line ratios for a single galaxy compared to the \cataldi calibrations in the left panel of \cref{fig:zdiag_branch_example}. All empirical calibrations, including the \cataldi calibrations, are valid within a certain metallicity regime, decided by the availability of data, which is denoted by thick solid lines in \cref{fig:zdiag_branch_example}. Green and blue circles denote observed ratios where reliable solutions are available, i.e., they coincide with the regime where the diagnostic is valid, and smaller gray circles denote a solution with extrapolation of the \cataldi calibration, and therefore unreliable. The horizontal lines denote observed ratios that are beyond the \cataldi calibrations, and therefore no metallicity solution is possible. Following the approach of \citet{Pilyugin:2005aa}, we chose to perform the \cataldi calibrations twice---adopting the higher (green circles) and lower (blue circles) metallicity branch in turn. In either scenario, there is a set of bins whose observed line ratios correspond to only one solution ---the lower metallicity branch. For these bins, a choice of metallicity branch was neither possible nor necessary. Therefore, adopting the higher metallicity branch (top right panel of \cref{fig:zdiag_branch_example}) leads to two distinct metallicity populations, whereas adopting lower metallicity branch (bottom right panel of \cref{fig:zdiag_branch_example}) automatically results in a smooth metallicity profile. Thus, even with the same observed line ratio profile, the resultant metallicity profile can vary significantly based on the above choice.

\subsection{Observed ratios beyond the calibration range}
\label{sec:outside_model}

The red shaded region in \cref{fig:zdiag_branch_example} corresponds to a non-existent solution, because the observed ratio is not covered by the model. \cref{fig:line_ratio_hist} shows the distribution of our observed line ratios in all the spatial bins for all galaxies, as well as global measurements, in the context of the different ranges of observed ratio, with the same colors as in \cref{fig:zdiag_branch_example}. Each bin of the histogram denotes the average S/N of all observed ratios in that bin. A significant fraction of the observed ratios (including integrated ratios) happen to lie outside the model coverage. Moreover, the observed ratios in this regime have a higher S/N, and therefore higher diagnostic power, making it even more challenging to derive metallicities.

We assigned the turnover metallicity value (vertical dashed line in \cref{fig:zdiag_branch_example}) to all bins (and integrated values) in the green region of \cref{fig:line_ratio_hist}, because that is the nearest metallicity from the observed ratio to the models. This can lead to an apparent `accumulation' of bins at a certain metallicity\, as seen in \cref{fig:zdiag_comp}. While not ideal, such ``fixed'' metallicity value is the closest to the truth that we can achieve with existing models.

Despite recent advances \citep{Cataldi:2025aa, Peluso:2025aa}, we do not yet have a clear choice for a robust metallicity diagnostic that self-consistently applies to a wide range of metallicity values seen here ($7<$ \logOH $<9$). Moreover, most of the above diagnostics were calibrated for global metallicities \citep{Kewley:2013ab, Dors:2020ab, Cataldi:2025aa} rather than spatially resolved, and therefore have not been tested for the wide range of line ratios observed at high-$z$. The ``excess'' line ratio presented here have been observed in previous intermediate-to-high redshift studies too \citep[e.g.,][]{Tollerud:2010aa, Florian:2021aa, Nakajima:2023aa}. Our analysis further confirms this issue and highlights the need for photoionization models that are able to adequately reproduce all observed line ratios at high-$z$. Therefore, in the absence of reliable ways to resolve the metallicity-branch degeneracy, and in the event of unreproducible observed ratios, metallicities estimated using strong emission line (SEL) diagnostics should be treated with caution.

\subsection{Metallicity diagnostics comparisons}
\label{sec:zcomp}

\begin{figure*}
    \centering
    \includegraphics[width=0.495\linewidth]{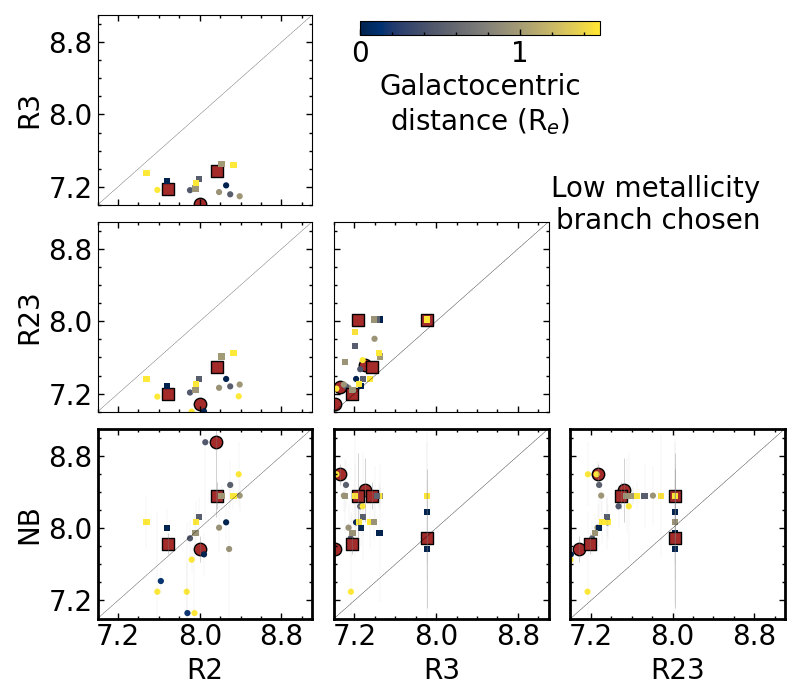}
    \includegraphics[width=0.495\linewidth]{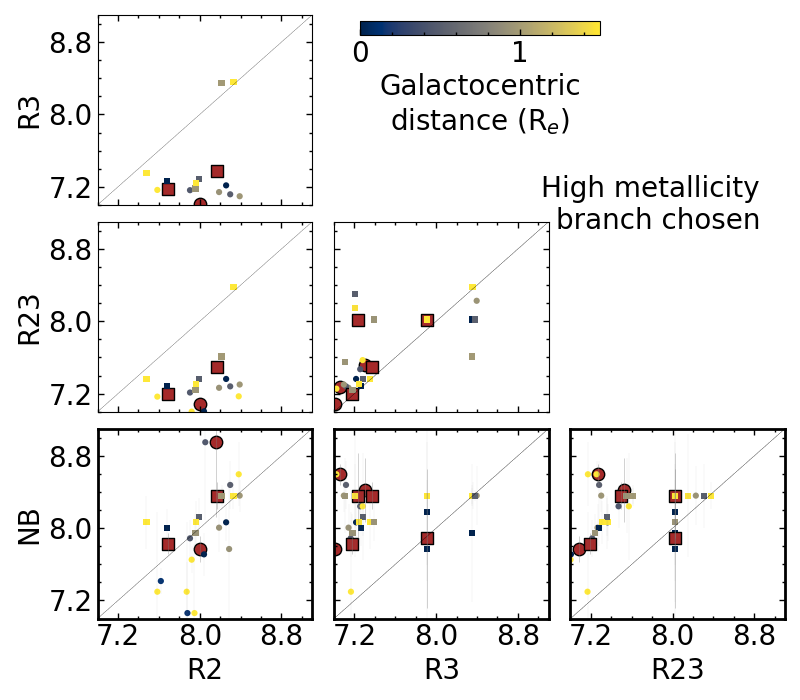}
    \caption{Comparing metallicities from different diagnostics---for both individual spatial bins (small symbols) as well as integrated measurements (large symbols). PASSAGE and GLASS galaxies are distinguished by using circle vs square markers, respectively. Each point is color-coded by the galactocentric distance of the corresponding bin. The \textit{left} and \textit{right} groups of plots correspond to the choice of lower and higher metallicity branches, respectively (see \cref{sec:double_valued}). The larger brown symbols denote integrated metallicities. The dashed diagonal lines represent the 1:1 scenario. The bottom rows and the left columns correspond to \texttt{NebulaBayes} (NB) and R2 metallicities, respectively, where the choice of branch is irrelevant. The NB rows are highlighted with thicker borders to denote that NB is our primary choice of metallicity diagnostic for all analysis hereafter. The `accumulation' of bins at certain metallicities, manifesting as `stripes' is discussed in \cref{sec:outside_model}.}
    \label{fig:zdiag_comp}
\end{figure*}

\cref{fig:zdiag_comp} shows a comparison of the spatially resolved metallicities obtained using the R2, R3, and R23 \cataldi calibrations as well as \texttt{NebulaBayes}. We performed the \cataldi diagnostics twice---by adopting the lower (left panel of \cref{fig:zdiag_comp}), and the higher metallicity branch (right panel), whenever double-valued solutions were available. However, given only a small fraction of our observed line ratios lay in the two-solution regime (white band; \cref{fig:line_ratio_hist}), the choice of metallicity branch had no impact on most spatial bins\footnote{The choice of branch can still significantly impact metallicity gradient measurements, even if a few spatial bins switch between the lower and higher metallicity branches.}.

Generally, the R3 and R23 calibrations agree well with each other. The agreement is better at lower metallicities, where \oiii is dominant over \oii. However, both R3 and R23 systematically yield significantly lower metallicities ($\sim1$ dex) than the R2 calibration. It is well known that metallicities derived with different calibrations can lead to discrepancies of $\lesssim0.8$ dex \citep[e.g.,][]{Kewley:2008aa, Lopez-Sanchez:2012aa}. However, the R2 metallicities agree best with \texttt{NebulaBayes}, albeit with a large scatter of $\sim0.5$ dex.

Upon adopting the higher metallicity solution for the \cataldi calibrations wherever available (right panel of \cref{fig:zdiag_comp}) the spatial bins get divided into two distinct populations---one where the higher metallicity solution was available and therefore adopted, and another where only the low metallicity solution existed. But given only a small fraction of our observed line ratios occupy the two-solution regime to begin with, choosing the higher metallicity branch only affects a very small number of spatial bins. For the emission line maps considered here, typically the centers exhibit higher line ratios than the outskirts. Therefore, the bins exhibit a bimodal metallicity distribution upon adopting the high metallicity branch.

Note that the `accumulation' of metallicities, manifesting as vertical or horizontal `stripes' in \cref{fig:zdiag_comp}, occurs due to the observed line fluxes being outside the model or calibration range. In \cref{sec:outside_model} we explained this for the SEL diagnostics. The Bayesian approach is not immune to this either. \texttt{NebulaBayes} weighs the observed fluxes against every model grid point and computes which model is most likely to reproduce them. Therefore, in cases where the observed flux ratios are outside the model grid, \texttt{NebulaBayes} treats the closest grid point as the best-fit model, leading to the `accumulation'. Owing to the utilisation of all available emission lines in \texttt{NebulaBayes}, along with its lack of dependence on user-defined choices such as the metallicity branch, \texttt{NebulaBayes} is a safer method to quantify spatially resolved metallicity. Therefore, we used only the \texttt{NebulaBayes} metallicities to measure gradients.

\subsection{Mass-metallicity gradient relation with strong-line diagnostics}
\label{sec:ap_mzgrad}

\begin{figure*}
    \centering
    \includegraphics[width=1\linewidth]{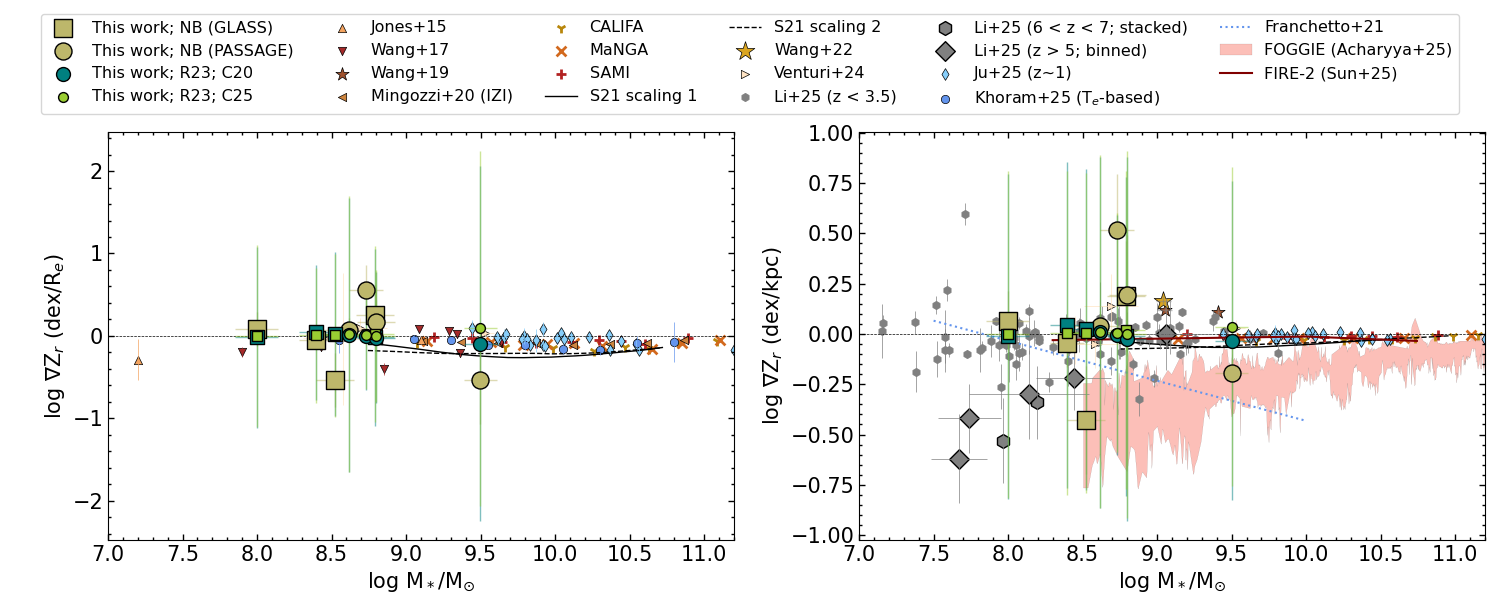}
    \caption{Same mass-metallicity gradient plot as \cref{fig:mzgrad}, but now including the gradients obtained using the R23 calibration (lower branch) from both \citet[][dark green]{Curti:2020ab} and \citet[][light green]{Cataldi:2025aa}.}
    \label{fig:mzgrad_r23}
\end{figure*}

\cref{fig:mzgrad_r23} shows that \curti and \cataldi diagnostics, which were deemed less reliable (see \cref{sec:zcomp}) lead to flatter gradients.

\section{Derivation of $t_{mix}$}
\label{sec:ap_tmix}

Here, we present the detailed derivation of the effective metal mixing timescale ($t_{mix}$) discussed in \cref{sec:disc}. We define the metallicity at a location as $Z(t) = \frac{\Sigma_m (t)}{\Sigma_g (t)}$, where $\Sigma_m$ and $\Sigma_g$ are the local metal mass and gas mass surface densities at a certain epoch $t$. We assume a linear radial profile, such that the metallicity gradient $\nabla Z = \frac{dZ}{dr}$ is constant.

Assuming all the stellar mass is formed out of the local gas mass, the rate of change of gas mass surface density ($\frac{d\Sigma_g}{dt}$) relates to the SFR surface density $\Sigma_{\rm SFR}$ as

\begin{equation}
\label{eq:kappa}
    \frac{d\Sigma_g}{dt} = -\Sigma_{\rm SFR}
\end{equation}

As with metallicity, we assume a smooth linear radial profile for $\Sigma_{\rm SFR}$. We let $f$ denote the metal mass transported per unit time. Combining metal transport and metal production terms, the rate of change in local metal budget is given by

\begin{equation}
    \frac{d\Sigma_m}{dt} = y \cdot \Sigma_{\rm SFR} - f = y \cdot \Sigma_{\rm SFR} - \frac{d\Sigma_m}{dL} \cdot \frac{dL}{dt}
\label{eq:dmdt}
\end{equation}

\noindent where $y$ is the metal yield, assumed constant, $L$ is the physical scale of the star-forming region, and $\frac{dL}{dt}$ is the effective gas velocity over this scale. We can further simply 
\begin{equation}
    \frac{d\Sigma_m}{dL} \cdot \frac{dL}{dt} \sim \frac{\Sigma_m}{t_{mix}}
\label{eq:dmdl}
\end{equation}

\noindent where $t_{mix}$ is a characteristic mixing timescale. Here, $t_{mix}$ is not intended to be specifically linked to metal mixing due to transport, but rather an effective timescale of mixing of metals relative to gas, under the net influence of diffusion, outflows and inflows. Upon substituting \cref{eq:dmdl} in to \cref{eq:dmdt}, the latter becomes

\begin{equation}
    \frac{d\Sigma_m}{dt} = y \cdot \Sigma_{SFR} - \frac{\Sigma_m}{t_{mix}}
\end{equation}

\noindent Therefore, the rate of change of metallicity is given by:

\begin{align}
\frac{dZ}{dt} &= \frac{1}{\Sigma_g}\cdot \frac{d\Sigma_m}{dt} - \frac{Z}{\Sigma_g} \cdot \frac{d\Sigma_g}{dt} \\
&= \frac{1}{\Sigma_g}(y \cdot \Sigma_{\rm SFR} - \frac{\Sigma_m}{t_{mix}}) + \frac{Z}{\Sigma_g} \cdot \Sigma_{\rm SFR}
\end{align}

\noindent Assuming quasi-steady-state ($\frac{dZ}{dt} \approx 0$),

\begin{equation}
0 = \frac{1}{\Sigma_g}(y \cdot \Sigma_{\rm SFR} - \frac{\Sigma_m}{t_{mix}}) + \frac{Z}{\Sigma_g} \cdot \Sigma_{\rm SFR}
\end{equation}

\noindent Solving for $Z$, we get
\begin{equation}
Z = \frac{y t_{mix}\left(\frac{\Sigma_{\rm SFR}}{\Sigma_g}\right)}{1 - t_{mix}\left(\frac{\Sigma_{\rm SFR}}{\Sigma_g}\right)}
\end{equation}

\noindent By considering $\epsilon = t_{mix}\left(\frac{\Sigma_{\rm SFR}}{\Sigma_g}\right)$, the above expression simplifies to
\begin{equation}
\label{eq:Z}
Z = \frac{y\epsilon}{1 - \epsilon}
\end{equation}

\noindent Assuming local Kennicutt-Schmidt (KS) relation \citep{Kennicutt:1998ab}, we relate $\Sigma_{\rm SFR} = A \Sigma_g^\alpha$. This gives

\begin{equation}
\label{eq:eps}
    \epsilon = t_{mix}\cdot B\Sigma_{\rm SFR}^\beta
\end{equation}

\noindent where $B=A^{1/\alpha}$ and $\beta = 1 - 1/\alpha$. We rearrange the constants in this way for ease of subsequent analysis.

Differentiating \cref{eq:Z} with respect to $\Sigma_{\rm SFR}$
\begin{equation}
\label{eq:slope}
    \frac{dZ}{d\Sigma_{\rm SFR}} = \frac{dZ}{d\epsilon} \cdot \frac{d\epsilon}{d\Sigma_{\rm SFR}} = \frac{y}{(1 - \epsilon)^2} \cdot \frac{d\epsilon}{d\Sigma_{\rm SFR}}
\end{equation}

\noindent As $\frac{dZ}{d\Sigma_{\rm SFR}}$ is on linear scale, whereas the slopes we derived in \cref{fig:zsfr} are in fact $\frac{d \log Z}{d\log \Sigma_{\rm SFR}}$ slopes, we converted the former using
\begin{equation}
\label{eq:log_slope}
    \frac{d \log Z}{d\log \Sigma_{\rm SFR}} = \frac{\Sigma_{\rm SFR}}{Z} \cdot \frac{dZ}{d\Sigma_{\rm SFR}}
\end{equation}

\noindent Combining \cref{eq:slope} and \cref{eq:log_slope}
\begin{equation}
    \frac{d \log Z}{d\log \Sigma_{\rm SFR}} = \frac{\Sigma_{\rm SFR}}{Z} \cdot \frac{y}{(1 - \epsilon)^2} \cdot \frac{d\epsilon}{d\Sigma_{\rm SFR}}
\end{equation}

\noindent Substituting \cref{eq:Z} and \cref{eq:eps}, the above becomes
\begin{align}
    \frac{d \log Z}{d\log \Sigma_{\rm SFR}} &= \Sigma_{\rm SFR} \cdot \frac{(1 - \epsilon)}{y\epsilon} \cdot \frac{y}{(1 - \epsilon)^2} \cdot t_{mix}\,B\,\beta\,\Sigma_{\rm SFR}^{\beta -1} \\
    &= \frac{\beta\,t_{mix}\,B\,\Sigma_{\rm SFR}^\beta}{1 - \,t_{mix}\,B\,\Sigma_{\rm SFR}^\beta}
\end{align}

\noindent Finally,
\begin{equation}
\label{eq:ap_tmix}
    \boxed{\frac{d \log Z}{d\log \Sigma_{\rm SFR}} = \beta \frac{\epsilon}{1 - \epsilon}\quad \text{where} \quad \epsilon = t_{mix}B\Sigma_{\rm SFR}^\beta}
\end{equation}

\noindent Assuming $A=2.5 \times 10^{-4}$ \citep{Kennicutt:1998ab} and $\alpha=1.5$ \citep{Kennicutt:2021aa} from typical KS relations, we found $B=0.004$ and $\beta=0.333$. Plugging all these values in \cref{eq:ap_tmix}, we inferred $t_{mix}$, in order to investigate any residual effect of stellar mass on $t_{mix}$.

\end{appendix}

\end{document}